%% file: xaaapaper.tex
\documentclass{article}
\usepackage{url}
\usepackage{setspace}
\usepackage{bigints}
\usepackage[dvipsnames]{xcolor}
\input{kvmacros}

\usepackage{graphicx}
\usepackage{endnotes}
\usepackage[a4paper,portrait]{geometry}
\usepackage{roundbox}
\usepackage{verbatimbox}
\usepackage{float}
\newcommand{\presentation}[2]{#1 }  

\usepackage{oz2e}
\usepackage{newcom}
\usepackage{proof}

\input{xaaamaincom.tex}

\title{Formal Analysis of Reachability, Infection and Propagation Conditions in Mutation Testing}
\author{Seyed-Hassan Mirian-Hosseinabadi\\
Associate Professor,\\
 Department of Computer Engineering,\\
 Sharif University of Technology\\
Email: hmirian@sharif.edu}
\begin{document}
\input{xaaastprog.tex}
\maketitle



\input{zp0000}
\input{zp0100}

\input{zp0200}

\input{zp0300}
\input{zp0401}
\input{zp0402}
\input{zp0403}
\input{zp0404}
\input{zp0405}
\input{zp0406}
\input{zp0500}

\input{zp0800-bib}

 \end{document}

\include{zp0000}
\include{zp0100}
\include{zp0200}
\include{zp0300}
\include{zp0401}
\include{zp0402}
\include{zp0403}
\include{zp0404}
\include{zp0405}
\include{zp0500}
\include{zp0800-bib}

%% file: xaaamaincom.tex
\newcounter{chap}
\renewcommand{\blseq}{[}
\renewcommand{\brseq}{]}

\newcounter{xmpl}
\newcounter{ques}
\newcounter{rese}

\input{xcasemgfcom.tex}



\input{xaaapdfsprog.tex}


\input{xaaapdfsdefi.tex}

\input{xaaapdfssynt.tex}

\input{xaaapdfsabbr.tex}
\input{xaaapdfsspec.tex}
\input{xaaapdfsprop.tex}
\input{xaaapdfsinfrl.tex}
\input{xaaapdfsprof.tex}
\input{xaaapdfsllaw.tex}


%% file: xaaapdfsprog.tex
\newcommand{\pminab}{m:= a\compose
\ifst b < a \rightarrow m := b
\elsest b \geq a \rightarrow \skipst
\fist\compose
\returnst(m)
}

\newcommand{\pminabma}{\mutatedpart{m:= b}\compose
\ifst b < a \rightarrow m := b
\elsest b \geq a \rightarrow \skipst
\fist\compose
\returnst(m)
}

\newcommand{\pminabmb}{m:= a\compose
\ifst \mutatedpart{b < m} \rightarrow m := b
\elsest \mutatedpart{b \geq m} \rightarrow \skipst
\fist\compose
\returnst(m)
}

\newcommand{\piseven}{
\ifst a < 0 \rightarrow a := 0- a
\elsest a \geq 0  \rightarrow \skipst
\fist\compose\\
\ifst \floor{\frac{a}{2}} = \frac{a}{2} \rightarrow \returnst(\truest) 
\elsest \floor{\frac{a}{2}} \neq \frac{a}{2} \rightarrow \returnst(\falsest)
\fist
}

\newcommand{\pisevenm}{
\ifst a < 0 \rightarrow \mutatedpart{a := 0}
\elsest a \geq 0  \rightarrow \skipst
\fist\compose\\
\ifst \floor{\frac{a}{2}} = \frac{a}{2} \rightarrow \returnst(\truest) 
\elsest \floor{\frac{a}{2}} \neq \frac{a}{2} \rightarrow \returnst(\falsest)
\fist
}

\newcommand{\pchkdigbb}{
 s := 1 \compose\\
\ifst (b\leq 1 \lor b > 10 \lor a < 0)  \rightarrow \returnst(s) \\
\elsest (b > 1 \land b \leq 10 \land a \geq 0) \rightarrow \skipst \compose\\   
s:= 2 \compose r :=a \compose d:=0 \compose \\
\dost  ( r > 0 \land d < b) \rightarrow\\ 
 \t1              t := r\compose r := \floor{\frac{t}{10}} \compose d := t-r*10\\
\odst}
\newcommand{\pchkdigab}{
\returnst(s)}

\newcommand{\pchkdigstub}{
\ifst ((d  < b)  \land  (r = 0)) \rightarrow s :=3\\
\elsest ((d \geq b) \lor (r \neq 0)) \rightarrow \skipst}

\newcommand{\pchkdigstmb}{
\ifst \mutatedpart{((d  < b)  \lor (r = 0))} \rightarrow s :=3\\
\elsest \mutatedpart{((d \geq b) \land (r \neq 0))} \rightarrow \skipst}
\newcommand{\pchkdigba}{
 s := 1 \compose\\
\ifst (b\leq 1 \lor b > 10 \lor a < 0)  \rightarrow \returnst(s) \\
\elsest (b > 1 \land b \leq 10 \land a \geq 0) \rightarrow \skipst \compose\\   
s:= 2 \compose r :=a \compose d:=0 
}
\newcommand{\pchkdigaa}{
\ifst ((d  < b)  \land  (r = 0)) \rightarrow s :=3\\
\elsest ((d \geq b) \lor (r \neq 0)) \rightarrow \skipst \compose\\
\returnst(s)}

\newcommand{\pchkdigprjba}{
             t := r\compose r := \floor{ \frac{t}{10}}
}
\newcommand{\pchkdigprjaa}{
\nullst}
\newcommand{\pchkdigprjstua}{
 d := t-r*10
}
\newcommand{\pchkdigprjstma}{
 \mutatedpart{d := t+r*10}
}
\newcommand{\pchkdiggja}{
( r > 0 \land d < b)
}
\newcommand{\pchkdigstuan}{
\dost  \pchkdiggja \rightarrow\\ 
 \t1            \pchkdigprjba  \compose \pchkdigprjstua\\
\odst}
\newcommand{\pchkdigstua}{
\dost  \pchkdiggja \rightarrow\\ 
 \t1            \pchkdigprjba  \compose \pchkdigprjstua \compose \pchkdigprjaa\\
\odst}
\newcommand{\pchkdigstma}{
\dost \pchkdiggja  \rightarrow\\ 
 \t1            \pchkdigprjba  \compose \pchkdigprjstma \compose \pchkdigprjaa\\
\odst}

\newcommand{\pchkdigstuifa}{
\ifst  \pchkdiggja \rightarrow\\ 
 \t1            \pchkdigprjba  \compose \pchkdigprjstua \compose \pchkdigprjaa\\
 \elsest \lnot \pchkdiggja \rightarrow \skipst\\
\fist}

\newcommand{\pchkdigstuifan}{
\ifst  \pchkdiggja \rightarrow\\ 
 \t1            \pchkdigprjba  \compose \pchkdigprjstua \\
 \elsest \lnot \pchkdiggja \rightarrow \skipst\\
\fist}
\newcommand{\pchkdigstmifa}{
\ifst \pchkdiggja  \rightarrow\\ 
 \t1            \pchkdigprjba  \compose \pchkdigprjstma \compose \pchkdigprjaa\\
  \elsest \lnot \pchkdiggja \rightarrow \skipst\\
\fist}

\newcommand{\pchkdig}{
\pchkdigbb  \compose\\
\pchkdigstub \compose\\
\pchkdigab}

\newcommand{\pchkdigmb}{
\pchkdigbb  \compose\\
\pchkdigstmb \compose\\
\pchkdigab}

\newcommand{\pchkdigma}{
\pchkdigba  \compose\\
\pchkdigstma \compose\\
\pchkdigaa}

\newcommand{\psearchb}{
 i := 0 \compose\\
 l := b.length 
}
\newcommand{\psearcha}{
\returnst(0)}
\newcommand{\psearchprjb}{
           \nullst 
}
\newcommand{\psearchprja}{
i:= i+1}
\newcommand{\psearchprjstu}{
\begin{array}{l}
 \ifst  (x=b~i) \rightarrow \returnst(1)\\
 \elsest  (x\neq b~i) \rightarrow \skipst\\
 \fist
  \end{array}
}
\newcommand{\psearchprjstm}{
\begin{array}{l}
  \ifst  \mutatedpart{(x\leq b~i)} \rightarrow \returnst(1) \\
   \elsest \mutatedpart{(x > b~i)} \rightarrow \skipst \\
  \fist
  \end{array}
}
\newcommand{\psearchgj}{
( i < l)
}

\newcommand{\psearchstu}{
\dost  \psearchgj \rightarrow\\ 
 \t1            \psearchprjb  \compose\\
 \t1  \psearchprjstu \compose\\
 \t1  \psearchprja\\
\odst}
\newcommand{\psearchstm}{
\dost \psearchgj  \rightarrow\\ 
 \t1            \psearchprjb  \compose \\
 \t1 \psearchprjstm \compose\\
 \t1  \psearchprja\\
\odst}

\newcommand{\psearchstuif}{
\ifst  \psearchgj \rightarrow\\ 
 \t2            \psearchprjb  \compose\\
 \t2  \psearchprjstu \compose\\
 \t2  \psearchprja\\
 \elsest \lnot\psearchgj \rightarrow \skipst\\ 
\fist}
\newcommand{\psearchstmif}{
\ifst \psearchgj  \rightarrow\\ 
 \t2            \psearchprjb  \compose \\
 \t2 \psearchprjstm \compose \\
 \t2 \psearchprja\\
  \elsest \lnot \psearchgj \rightarrow \skipst\\ 
\fist}

\newcommand{\psearch}{
\psearchb  \compose\\
\psearchstu \compose\\
\psearcha}

\newcommand{\psearchm}{
\psearchb  \compose\\
\psearchstm \compose\\
\psearcha}



%% file: xaaapdfsdefi.tex
\newcommand{\daswp}{\wpre{w:=E,\Gpost} \defs \Gpost\rN{w}{E}}

\newcommand{\dasrc}{\reachc{w:=E} \defs \truest}

\newcommand{\dskwp}{\wpre{\skipst,\Gpost} \defs \Gpost}

\newcommand{\dskrc}{\reachc{\skipst} \defs \truest}

\newcommand{\dabwp}{\wpre{\abortst,\Gpost} \defs \falsest}

\newcommand{\dabrc}{\reachc{\abortst} \defs \falsest}

\newcommand{\dscwp}{\wpre{\PRo\compose \PRt, \Gpost} \defs \wpre{\PRo,\wpre{\PRt,\Gpost}}}

\newcommand{\dscrc}{\reachc{\PRo\compose \PRt} \defs  \reachc{\PRo} \land \reachc{\PRt}}

\newcommand{\difwp}{
 \begin{array}{l}
\wpre{\ifst (\elsest i @ G_i \rightarrow \PRi) \fist, \Gpost} \defs\\
\t1  (\lor_i @ G_i) \land ( \land_i @ G_i \implies \wpre{\PRi,\Gpost})
\end{array}}

\newcommand{\difrc}{
 \begin{array}{l}
\reachc{\ifst (\elsest i @ \Gi \rightarrow \PRi) \fist} \defs ( \lor_i @  \Gi \land \reachc{\PRi})
\end{array}}

\newcommand{\ddowpDa}{
 \begin{array}{l}
\DO \defs \dost (\elsest i . \Gdoi \rightarrow \PRi) \odst ~\\
\IF \defs \ifst (\elsest i . \Gdoi \rightarrow \PRi) \fist ~\\
\Gdo \defs \lor_{i} \Gdoi\\ 
\DO \defs \dost \Gdo \rightarrow \IF \odst
\end{array}
}
\newcommand{\ddowpDba}{
\begin{itemize}
\item{$\Ido$,} is a predicate that is true before and after each iteration of a loop, which is called {\it Invariant}.
\item{$\Vdo$,}  is an integer function of program variables that is an upper bound on number of iterations still to be performed. Each iteration of the loop decreases $\Vdo$ by at least $1$ and, as long as execution of the loop has not terminated, $\Vdo$ is bounded below by $0$. Hence, the loop must terminate. Function $\Vdo$ has been called a {\it variant} function.
\item{$\Vardo$,}  is the list of variables modified within the loop.
\item{$\Varfreshdo$,}  is a fresh variable.
\end{itemize}}
\newcommand{\ddowpDbb}{
\begin{itemize}
\item{$\DHp{0}{\Gpost}$,} represents the set of states in which execution of $\DO$ terminates in $0$ iteration with $\Gpost$ true. For zero iteration all guards must be $\falsest$.
\item{$\DHp{k}{\Gpost}~for~k>0$,} represents the set of states in which execution of $\DO$ terminates in $k$ or fewer iterations with $\Gpost$ true. The definition is recursive. One case is that $\DO$ terminates in $0$ iteration, in which case $\DHp{0}{\Gpost}$ is true. The other case is that at least one iteration is performed. Thus, $\Gdo$ must initially be true and the iteration consists of executing a corresponding $\IF$. This execution of $\IF$ must terminate in a state in which the loop will iterate $k-1$ or fewer times.
\item{$\wpre{\DO, \Gpost}$,} represents the set of states in which execution of $\DO$ terminates in a bounded number of iterations with $\Gpost$ true. That is, initially there must be some $k$ such that at most $k$ iterations will be performed.
\end{itemize}}

\newcommand{\ddowpDc}{
Loop Well-Definedness Conditions($\LWDC$):
\begin{displaymath}
\forall \Vardo @ (\forall i:\nat | 1 \leq i \leq n @   \\
\t2 (\Ido \land \Gdoi \implies \wpre{\PRi,\Ido}) \land  \\
\t2  (\Ido \land \Gdoi \implies  \wpre{\Varfreshdo:=\Vdo \compose \PRi, \Vdo < \Varfreshdo}))  \land \\
\t1 (\Ido \land \Gdo \implies \Vdo >0)\\
\end{displaymath}
}

\newcommand{\ddowpDce}{
Loop Well-Definedness Conditions($\LWDC$) obligations have the following meanings :
\begin{itemize}
 \item $(\Ido \land \Gdoi \implies \wpre{\PRi,\Ido})$  \\
 The loop invariant $\Ido$ is maintained by the loop body, provided the loop is entered: this acts like an induction step in a proof that $\Ido$ is always $\truest$ at the beginng of each loop body execution, and at termination of the loop.
 \item $(\Ido \land \Gdoi \implies  \wpre{\Varfreshdo:=\Vdo \compose \PRi, \Vdo < \Varfreshdo}))$ \\
The variant $\Vdo$ is always strictly decreased by each loop body execution.
\item $(\Ido \land \Gdo \implies \Vdo >0)$\\
During loop execution, the variant $\Vdo$  is always greater than zero: it will usually be an expression involving some of the variables modified within the loop.
\end{itemize}
}

\newcommand{\ddowpDdh}{
\begin{array}{l}
\DHp{0}{\Gpost} = (\Gpost \land \lnot \Gdo) \\
\DHp{k}{\Gpost} = (\wpre{\IF, \DHp{k-1}{\Gpost}}) \\
\wpre{\DO, \Gpost}= \exists k:\nat @ \DHp{k}{\Gpost}\\
\end{array}
 }

\newcommand{\ddorcMb}{
\begin{itemize}
\item{$\MRp{0}$,} represents the condition  which execution of $\DO$ generates in $0$ iteration. For zero iteration all guards must be $\falsest$.
\item{$\MRp{k}~for~k>0$,} represents the condition which execution of $\DO$ generates in $k$ or fewer iterations. The definition is recursive. One case is that $\DO$ terminates in $0$ iteration, in which case $\MRp{0}$ is true. The other case is that at least one iteration is performed. Thus, $\Gdo$ must initially be true and the iteration consists of executing a corresponding $\IF$. This execution of $\IF$ generates some rechability conditions and must terminate in a state in which the loop will iterate $k-1$ or fewer times.
\item{$\reachc{\DO}$,} represents the conditions  which execution of $\DO$ generates in a bounded number of iterations. That is, initially there must be some $k$ such that at most $k$ iterations will be performed.
\end{itemize}}

\newcommand{\ddorcMd}{
\begin{array}{l}
\MRp{0} = (\lnot \Gdo) \\
\MRp{k} = (\reachc{\IF} \land  \MRp{k-1})\\
\reachc{\DO}= \exists k:\nat @ \MRp{k}\\
\end{array}
 }

\newcommand{\drtwp}{\wpre{\returnst(E),\Gpost} \defs \Gpost\rN{\Gpn}{E}}

\newcommand{\drtrc}{\reachc{\returnst(E)} \defs \falsest}

\newcommand{\dnlwp}{\wpre{\nullst,\Gpost} \defs \Gpost}

\newcommand{\dnlrc}{\reachc{\nullst} \defs \truest}



%% file: xaaapdfssynt.tex
\newcommand{\yasst}{w:=E}

\newcommand{\yskst}{\skipst}

\newcommand{\yrtst}{\returnst(E)}

\newcommand{\yscst}{P_1 \compose P_2}

\newcommand{\ynlst}{\nullst}

\newcommand{\yabst}{\abortst}

\newcommand{\ymast}{w_1,\ldots,w_n := E_1,\ldots,E_n}

\newcommand{\yifstv}{\ifst G_1 \rightarrow prog_1\\
\vdots\\
\elsest G_j \rightarrow prog_j\\
\vdots\\
\elsest G_n \rightarrow prog_n\\
\fist}
\newcommand{\yifsth}{
\ifst (\elsest i . G_i \rightarrow prog_i) \fist}

\newcommand{\ydostv}{\dost G_1 \rightarrow prog_1\\
\vdots\\
\elsest G_j \rightarrow prog_j\\
\vdots\\
\elsest G_n \rightarrow prog_n\\
\odst}
\newcommand{\ydosth}{
\dost (\elsest i . G_i \rightarrow prog_i) \odst}

%% file: xaaapdfsllaw.tex
\newcommand{\lAonA}{(\Gpa \lor \lnot \Gpa) \equiv \truest}

\newcommand{\lAiB}{(\Gpa \implies \Gpb) \equiv (\lnot \Gpa \lor \Gpb)}

\newcommand{\lAoBiC}{(\Gpa \lor \Gpb) \implies \Gpc \equiv (\Gpa \implies \Gpc) \land (\Gpb \implies \Gpc)}

\newcommand{\lAaBiC}{(\Gpa \land \Gpb) \implies \Gpc \equiv (\Gpa \implies \Gpc) \lor (\Gpb \implies \Gpc)}

\newcommand{\lAiBanAiC}{(\Gpa \implies \Gpb) \land (\lnot \Gpa \implies \Gpc) \equiv (\Gpa \land \Gpb) \lor (\lnot \Gpa \land \Gpc)}

\newcommand{\lNcP}{(\nullst \compose prog) = prog}

\newcommand{\laneb}{(a \neq b) = (a < b) \lor (a >b)}

\newcommand{\laleb}{(a \leq b) = (a = b) \lor (a <b)}

%% file: xaaastprog.tex
\begin{myverbbox}{\pminabvb}
1. int Min (int A, int B)
2. {
3.        int minVal;
4.        minVal = A;
4'.       minVal = B; // Mutant 1
5.        if (B < A)
5'.       if (B < minval) // Mutant 2
6.        {
7.             minVal = B; 
8.         }
9.         return (minVal);
10. } // end Min
\end{myverbbox}
\begin{myverbbox}{\pisevenvb}
1     boolean isEven (int X)
2     {
3          if (X < 0)
4               X = 0 - X;
4'              X = 0;  // Mutant 
5           if (double) (X/2) == ((double) X) / 2.0
6               return (true);
7           else
8               return (false);
9     }
\end{myverbbox}
\begin{myverbbox}{\pchkdigvb}
1.  Procedure Chkdig (b:int,a:int)  return(s:int)
 2.  { s := 1;
 3.    if b <= 1 OR b > 10  OR a  < 0   then return(s);
 4.    s:= 2; r :=a; d:=0;
 5.    do while ( r > 0 AND d < b)
 6.           t := r;
 7.           r := INT(t/10);
 8.           d := t-r*10
 8'.          d := t+r*10   // Mutant 1
 9.    end;
10.    if d  < b  AND  r = 0 then s :=3 
10'.   if d  < b  OR   r = 0 then s :=3  // Mutant 2
11.   return(s); 
12.   }
\end{myverbbox}

\begin{myverbbox}{\psearchvb}
// If b is null throw NullPointerException
// else if x is in b then return 1  
// else return 0
1. public static int Search(int[] b, int x )
2. {
3.     for (int i=0; i < b.length; i++) 
4.    {
5.          if x == b[i] then return 1;
5'.         if x <= b[i] then return 1;
6.     }
7.    return 0;
8. }
\end{myverbbox}

%% file: zp0000.tex
\begin{abstract}
Finding test cases to kill the alive mutants in Mutation testing needs to calculate the Reachability, Infection and Propagation(RIP) conditions and the full test specification.
In this paper, a formal approach to calculate RIP conditions is proposed. The Dijkestra's weakest precondition predicate transformer ($\wpre{\_,\_}$)  is used to calculate infection and propagation conditions. The $\reachc{\_}$ function is defined to calculate the  reachability conditions generated by each statement.
Four programs and their mutants are examined as running examples and as  case studies to show the applicability of the method.
\end{abstract}  

%% file: zp0100.tex
\newcommand{\lastd}[1]{(#1-\floor{ \frac{#1}{10}}*10)}
\newcommand{\alld}[2]{(#1)_{#2}}
\setcounter{chap}{1}
\section{Introduction}
Mutation testing is the widely used method of a category of testing methods called Syntax Based Testing. This is a method to assess the quality of test cases by killing mutants, and at the same time,  to generate new test cases to kill those mutants which are alive. These new test cases can find faults in the program under test if there are still any, and can increase the level of confidence to the under test program. For those mutants which are not killed by the original set of test cases, the tester should find new test cases by the help of RIPR analysis\cite{testbook}.  To apply formal method to RIPR analysis, the programs are translated to  guarded commands  syntax,  and the weakest precondition predicate transformer and the reachability condition generator (defined in Section \ref{propsec}) are used\cite{Dijkestra}. In the rest of this section, first,  the Mutation Testing and RIPR analysis are discussed. Then, the syntax of Dijkestra Guarded Command Language is given by the help of BNF notations, and finally,   the Dijkestra's weakest precondition predicate transformer is discussed. In the rest of the paper, the naming and abbreviation standards defined in Table \ref{natable} are used:
\renewcommand{\ST}{st}
\newcommand{\MD}{module}
\newcommand{\mutated}[1]{#1^{m}}
\newcommand{\mutateda}[1]{#1^{m_{1}}}
\newcommand{\mutatedb}[1]{#1^{m_{2}}}
\newcommand{\undertest}[1]{#1^{u}}
\newcommand{\stmod}[1]{#1_{s}}
\newcommand{\gdmod}[1]{#1_{g}}
\newcommand{\before}[1]{#1^{b}}
\newcommand{\after}[1]{#1^{a}}

\newcommand{\MDsu}{\stmod{\undertest{\MD}}}
\newcommand{\MDsm}{\stmod{\mutated{\MD}}}
\newcommand{\MDgu}{\gdmod{\undertest{\MD}}}
\newcommand{\MDgm}{\gdmod{\mutated{\MD}}}
\newcommand{\STz}{\ST_{0}}
\newcommand{\STj}{\ST_{j}}
\newcommand{\STi}{\ST_{i}}
\newcommand{\STn}{\ST_{n}}
\newcommand{\PRu}{\undertest{\PRg}}
\newcommand{\PRm}{\mutated{\PRg}}
\newcommand{\STu}{\undertest{\ST}}
\newcommand{\PRjb}{\before{\PRj}}
\newcommand{\Go}{G_{1}}
\newcommand{\Gj}{G_{j}}
\newcommand{\Gi}{G_{i}}
\newcommand{\Gn}{G_{n}}
\newcommand{\STju}{\undertest{\STj}}
\newcommand{\PRja}{\after{\PRj}}
\newcommand{\STm}{\mutated{\ST}}
\newcommand{\STjm}{\mutated{\STj}}
\newcommand{\PRb}{\before{\PRg}}
\newcommand{\PRa}{\after{\PRg}}
\newcommand{\PRju}{ \PRjb \compose  \STju \compose  \PRja }
\newcommand{\PRjm}{\PRjb \compose  \STjm \compose  \PRja}

\begin{table}[h!]
\caption{Naming and Abbreviation Standards}
\centering
$
\begin{array}{|c|l|l|}
\hline
\mbox{Abbreviation} & \mbox{Long} & \mbox{Comment}\\
\hline
\ST & \mbox{Statement} & \mbox{Any statement of the program}\\
\PRg & \mbox{Program} & \mbox{An ordered list of statement}\\
\MD & \mbox{Module} & \mbox{Any ordered list of programs}\\
\stmod{\_} & \mbox{statement  modification} & \mbox{statement modification type}\\
\gdmod{\_} & \mbox{guarded command  modification} & \mbox{guarded command modification type}\\
\undertest{\_} & \mbox{under  test} & \mbox{part of the code which is under test}\\
\mutated{\_} & \mbox{mutant} & \mbox{part of the code which is mutated} \\
\after{\_} &  \mbox{after} & \mbox{part of the code which is after the mutated statement}\\
\before{\_} & \mbox{before} & \mbox{part of the code which is before the mutated statement} \\
\uparrow & \mbox{above} & \mbox{above calculations} \\
\lastd{a} & \mbox{last digit} & \mbox{last digit of $a$}\\
\alld{a}{b} & \mbox{all digits in base} & \mbox{all digits of $a$ are in base $b$}\\
\hline
\end{array}
$
\label{natable}
\end{table}
\subsection{Mutation Testing and RIPR Analysis}
\label{ssmtripra}
Mutation Testing is the original and most widely known application of Syntax-Based Testing.
In Syntax-Based Testing  a syntactic description of some data of the under test program is used  to define the test cases.
In Mutation Testing, the grammar of the  programming language is used as the syntactic description,  the program under test ($P$). and the mutants ($M$), are valid strings of this grammar.

 In Mutation Testing, Mutation Operators are selected to apply  to the source program one at a time for each applicable piece of the source code. 
 A mutant is the result of applying one mutation operator to the under test program.
The goals of Mutation Testing are:
\begin{itemize} 
\item to assess the quality of the test cases which should be robust enough to kill mutants. 
\item to find test cases (using RIPR analysis) to kill alive mutants.
\end{itemize}

The RIPR Analysis forces us to find following conditions:
\begin{itemize}
\item Reachability: The conditions need the faulty statement to be reached,
\item Infection: The conditions need the faulty statement to result in an incorrect state,
\item Propagation: The conditions need the incorrect state propagates to incorrect output,
\item Revealability: The conditions need the tester observe part of the incorrect output (not discussed in this paper).
\end{itemize}

When the under test program  is mutated to create mutants, the hope is to exhibit different behavior from the original program. Killing mutants is defined as follow:
\begin{Definition}
Killing Mutants: Given a mutant $m \in M$ for an under test program  $P$ and a test $t$, $t$ is said to kill $m$ if and only if the output of $t$ on $P$ is different from the output of $t$ on $m$.
\end{Definition}

If all the mutants cannot be killed by the existing set of tests the testers can keep adding tests by the help of RIPR analysis  until all mutants have been killed.
There are two approaches of killing mutants, namely, strongly killing mutants and weakly killing mutants.

\begin{Definition}
 Strongly Killing Mutants:
   Given a mutant $m\in M$ for a program $P$ and a test $t$, $t$ is said to strongly kill $m$ if and only if the output of $t$ on $P$ is different from the output of $t$ on $m$.
\end{Definition}
Strongly killing satisfies reachability and infection and  propagation conditions.

 \begin{Definition}
 Weakly Killing Mutants:
   Given a mutant $m \in M$ that modifies a location $L$ in a program $P$,  and a test $t$, $t$ is said to weakly kill $m$ if and only if the state of the execution of $P$ on $t$ is different from the state of the execution of $m$   on $t$ immediately after $L$.
   \end{Definition}
Weakly killing satisfies reachability and infection conditions, but not propagation condition.

In the rest of this paper, four programs and their mutants, two from Introduction to Software Testing book\cite{testbook} and two from my own examination  questions, are used in different examples. These examples are formally discussed in Section \ref{scase}.
\begin{EExample}
\label{exampmin}
Calculate the Reachability, Infection, Propagation and Full Test Specification conditions for two mutants of {\tt Min} program\cite{testbook}. The original code in which the mutated statements are marked by prime is as follow:

\pminabvb

Solution: No conditional statements are before mutated statement of Mutant 1 of {\tt Min} program,
thus the reachability condition is $\truest$. In order to infect, $({\tt minVal} = {\tt A}) \neq  ({\tt minVal} = {\tt B})$, 
 which is simplified 
as $({\tt A} \neq {\tt B})$. To propagate, the value of {\tt B} must not be the minimum value. It means $({\tt B} < {\tt A}) = \falsest$. The complete test
specification to kill this mutant strongly is:
\begin{argue}
Reachability: \truest\\
Infection: {\tt A} \neq {\tt B}\\
Propagation: {\tt B} < {\tt A} = \falsest\\
Full~ Test~ Specification: \truest \land ({\tt A} \neq {\tt B}) \land ({\tt B} < {\tt A} = \falsest)\\
\t3 ({\tt A} \neq {\tt B}) \land ({\tt B} \geq {\tt A})\\
\t3 {\tt B} > {\tt A}
\end{argue}

The reachability condition for mutated statement in  Mutant 2 of {\tt Min} is
$\truest$. The infection condition is $({\tt B} < {\tt A}) \neq ({\tt B} < {\tt minVal})$. 
 
\begin{argue}
Reachability: \truest\\
Infection: ({\tt B} < {\tt A}) \neq ({\tt B} < {\tt minVal})= & (Adding ${\tt minVal} = {\tt A}$)\\
\t3 (({\tt B} < {\tt A}) \neq ({\tt B} < {\tt minVal})) \land ({\tt minVal} = {\tt A})= & ($\neq$ properties)\\
\t3 ((({\tt B} < {\tt A}) \land ({\tt B} \geq {\tt minVal})) \lor (({\tt B} \geq {\tt A}) \land ({\tt B} < {\tt minVal})) \land ({\tt minVal} = {\tt A})=\\
&(Rearranging)\\
\t3 ((({\tt A} > {\tt B}) \land ({\tt B} \geq {\tt minVal})) \lor (({\tt A} \leq {\tt B}) \land ({\tt B} < {\tt minVal})) \land ({\tt minVal} = {\tt A})=\\
 &(Transitivity)\\
\t3 (({\tt A} > {\tt minVal}) \lor ({\tt A}  < {\tt minVal})) \land ({\tt minVal} = {\tt A})=& ($\neq$ properties)\\
\t3 ({\tt A} \neq {\tt minVal}) \land ({\tt minVal} = {\tt A})=\\
\t3 \falsest
\end{argue}
{\it The contradiction means that infection condition cannot be satisfied, as a result, the  mutant is equivalent to the original program.}
\end{EExample}
\begin{EExample}
\label{exampiseven}
Calculate the Reachability, Infection, Propagation and Full Test Specification conditions for the  mutant of {\tt isEven} program\cite{testbook}. The original code in which the mutated statements are marked by prime is as follow:

\pisevenvb

Solution: The reachability condition for the Mutant of {\tt isEven} program is $({\tt X} < 0)$ and the infection
condition is $({\tt X} =0- {\tt X}) \neq ({\tt X} = 0)$.  The propagation
condition for this mutant is  $(odd {\tt X})$. Thus, to kill this mutant strongly, it is  required:
\begin{argue}
Reachability: ({\tt X} < 0)\\
Infection: ({\tt X} =0- {\tt X}) \neq ({\tt X} = 0) =\\
\t3 ({\tt X} \neq 0)\\
Propagation: odd({\tt X})\\
Full~ Test~ Specification: ({\tt X} < 0) \land ({\tt X} \neq 0) \land odd({\tt X})=\\
\t4({\tt X} < 0)  \land odd({\tt X})\\
\end{argue}
\end{EExample}
\begin{EExample}
\label{exampchkdig}
Calculate the Reachability, Infection, Propagation and Full Test Specification conditions for the  mutant of {\tt Chkdig} program. The original code in which the mutated statements are marked by prime is as follow:

\pchkdigvb

Solution: The reachability condition for Mutant 1 of {\tt Chkdig} program  is $\lnot ({\tt b} \leq 1 \lor {\tt b} > 10  \lor  {\tt a}  < 0) \land ({\tt r} > 0 \land  {\tt d} < {\tt b})$  and the infection
condition is $({\tt d} := {\tt t} - {\tt r}*10) \neq ({\tt d} := {\tt t} + {\tt r}*10 )$. The propagation
condition is $\alld{a}{b}$. The complete test
specification to kill this mutant is:
\begin{argue}
Reachability: \lnot ({\tt b} \leq 1 \lor {\tt b} > 10  \lor  {\tt a}  < 0) \land ({\tt r} > 0 \land  {\tt d} < {\tt b})=\\
\t2 ({\tt b} > 1 \land {\tt b} \leq 10 \land {\tt a} \geq 0 \land {\tt a} > 0 \land  {\tt 0} < {\tt b})=\\
\t2 ({\tt b} > 1 \land {\tt b} \leq 10  \land {\tt a} > 0 )\\
Infection: ({\tt d} := {\tt t} - {\tt r}*10) \neq ({\tt d} := {\tt t} + {\tt r}*10 )=\\
\t2 {\tt r} > 0 = & \Argcalc{r=\floor{ \frac{a}{10}}}\\
\t2  {\tt a} \geq 10 \\
Propagation: \alld{a}{b}\\
Full~ Test~ Specification: {\tt b} > 1 \land {\tt b} \leq 10   \land {\tt a} > 0 \land {\tt a} \geq 10 \land \alld{a}{b}= \\
\t3 {\tt b} > 1 \land {\tt b} \leq 10  \land {\tt a} \geq 10 \land \alld{a}{b}\\
\end{argue}
The reachability condition for Mutant 2 of {\tt Chkdig} program is $\lnot ({\tt b} \leq 1 \lor {\tt b} > 10  \lor  {\tt a}  < 0) \land \lnot({\tt r} > 0 \land  {\tt d} < {\tt b})$  and the infection
condition is $({\tt d}  < {\tt b} \land  {\tt r} = 0) \neq ({\tt d}  < {\tt b}  \lor   {\tt r} = 0)$. The propagation
condition is $\truest$. The complete test
specification to kill this mutant is:
\begin{argue}
Reachability: \lnot ({\tt b} \leq 1 \lor {\tt b} > 10  \lor  {\tt a}  < 0) \land \lnot({\tt r} > 0 \land  {\tt d} < {\tt b})=\\
\t2 ({\tt b} > 1 \land {\tt b} \leq 10 \land {\tt a} \geq 0) \land ( {\tt r} \leq  0 \lor {\tt d} \geq {\tt b})\\
Infection: ({\tt d}  < {\tt b} \land  {\tt r} = 0) \neq ({\tt d}  < {\tt b}  \lor   {\tt r} = 0)=\\
\t2  {\tt a} > 0 \land \lastd{a} \geq {\tt b}\\
Propagation: \truest\\
Full~ Test~ Specification: (({\tt b} > 1 \land {\tt b} \leq 10 \land {\tt a} \geq 0) \land ( {\tt r} \leq  0 \lor {\tt d} \geq {\tt b})) \land ({\tt a} > 0 \land \lastd{a} \geq {\tt b})\\
\end{argue}
\end{EExample}
\subsection{Dijkestra Guarded Commands}

The syntax of Dijkestra Guarded Command  is given in BNF (Figure \ref{dgcfig}) with slightly modifications from the original one given in Dijkestra paper\cite{Dijkestra}.   

\begin{figure}[h]
\scriptsize
\roundbox{
\begin{minipage}{10cm}
\begin{syntax}
\langle program \rangle &::= & \langle statement~list \rangle \\
\langle statement~list\rangle & ::= & \langle statement \rangle |
\langle statement \rangle \compose \langle statement~list \rangle\\     
\langle statement \rangle & ::= & 
\langle alternative~construct \rangle | 
 \langle repetitive~construct \rangle | 
\langle assignment~statement \rangle | \\
                                     &  &      
\langle skip~statement \rangle | 
\langle return~statement \rangle | 
\langle abort~statement \rangle | \\         
                                    &  &      
\langle null~statement \rangle | 
 ``other~statements"\\ 
 \langle alternative~construct \rangle &::= & \ifst \langle guarded~command~set \rangle \fist\\     
 \langle repetitive~construct \rangle & ::= & \dost \langle guarded~command~set \rangle \odst\\  
 \langle guarded~command~set \rangle & ::= & \langle guarded~command \rangle | 
\langle guarded~command \rangle \elsest \langle guarded~command~set \rangle\\ 
 \langle guarded~ command \rangle & ::= &\langle guard \rangle \rightarrow \langle statement~list \rangle \\
\langle guard \rangle & ::= &  \langle boolean~expression \rangle\\ 
\langle assignment~statement \rangle & ::=& \langle identifier~list \rangle  := \langle expression~list \rangle\\
\langle skip~statement \rangle & ::=& \skipst\\
\langle return~statement \rangle & ::= & \returnst | \returnst(\langle expression \rangle)\\
\langle abort~statement \rangle & ::=& \abortst\\
\langle null~statement \rangle & ::=&     \nullst \\
\langle identifier~list \rangle  & ::= & ``w"\\
\langle expression~list \rangle & ::= & ``E"\\    
\end{syntax}
\end{minipage}
}
\normalsize
\caption{\label{dgcfig}
BNF description of Dijkestra Guarded Commands syntax}
\end{figure}

Consider the following notes which explain the notations used in the grammar.
\begin{itemize}
\item ``other statements" is indicating, other statements such as procedure calls. 
\item $``w"$ is any list of identifiers separated by commas.
\item $``E"$ is any list of expressions separated by commas.
\item The semicolon ($\compose$) in this grammar  is the sequential composition operator. 
\item The sequential composed statements  are executed successively in the 
order from left to right. 
\item When  a guard is true its  guarded list can be executed.
\item  The separator $\elsest$ is used to separate guarded commands of a guarded command list.
\item The order of  guarded commands of a set is not important. 
\item In the alternative 
construct, $\ifst \ldots \fist$,   
one of the  guarded list with a true guard will be executed. If none of the 
guards is true, the program will abort.
\item The repetitive construct, $\dost  \ldots \odst$,  will 
only terminate in a state in which none of the guards 
are true. 
\item Assignment, skip, return, abort, and null  statements  are added to the original grammar given by Dijkestra \cite{Dijkestra} to be precise about the syntax of these statements. The semantics of these statements are as usual. 
\item $\nullst$ is added to be used as a program which has no statements. 
\item  $\returnst(E)$ is treated as an assignment of $E$ to the program name $\Gpn$. 
\end{itemize}

Table \ref{ggctab} shows  generic examples of Dijkestra Guarded Commands statements.

\begin{table}[h]
\small
\centering
\caption{Generic examples of Dijkestra Guarded Commands statements}
$
\begin{array}{|l|l|}
\hline
\mbox{Construct Name} & \mbox{Example}\\
\hline
Assignment & \yasst\\
\hline
Multiple ~ Assignment & \ymast\\
\hline
Skip & \yskst\\
\hline
Sequential~ Composition & \yscst\\
\hline
Alternative~ Construct(Vertical) & \begin{array}{l}\yifstv \end{array}\\
\hline
Alternative~ Construct(Horizontal) & \begin{array}{l}\yifsth \end{array}\\
\hline
Repetitive~ Construct(Vertical) & \begin{array}{l} \ydostv \end{array}\\
\hline
Repetitive~ Construct(Horizontal) & \begin{array}{l} \ydosth \end{array}\\
\hline
Return & \yrtst\\
\hline
Abort & \yabst\\
\hline
Null & \ynlst\\
\hline
\end{array}
$
\label{ggctab}
\normalsize
\end{table}

 \newcommand{\AMD}{Min}
\newcommand{\AMDu}{\undertest{\AMD}}
\newcommand{\AMDma}{\mutateda{\AMD}}
\newcommand{\AMDmb}{\mutatedb{\AMD}}

In the following examples, the underlined statements are the mutated statements. 
\begin{EExample}
\label{exampmingc}
Abbreviating {\tt minVal} by {\tt m}, {\tt A} by {\tt a} and {\tt B} by {\tt b}, for the program and its mutants given in Example \ref{exampmin},
 \, the translation of this program and its mutated versions to Dijkestra Guarded Commands are:
\[\AMDu\defs \pminab \]
\[\AMDma \defs\pminabma \]
\[\AMDmb\defs\pminabmb \]
\end{EExample}

 \renewcommand{\AMD}{isEven}
\renewcommand{\AMDu}{\undertest{\AMD}}
\newcommand{\AMDm}{\mutated{\AMD}}

\begin{EExample}
\label{exampisevengc}
Abbreviating {\tt X} by {\tt a}, for the program and its mutant given in Example \ref{exampiseven},
 \, the translation of this program and its mutated version to Dijkestra Guarded Commands are:

\[\AMDu\defs \begin{array}{l} \piseven \end{array} \]
\[\AMDm \defs\begin{array}{l} \pisevenm \end{array} \]
\end{EExample}
 \renewcommand{\AMD}{Chkdig}
\renewcommand{\AMDu}{\undertest{\AMD}}
\renewcommand{\AMDma}{\mutateda{\AMD}}
\renewcommand{\AMDmb}{\mutatedb{\AMD}}
\begin{EExample}
\label{exampchkdiggc}
The translation of the program and its mutated versions given in Example \ref{exampchkdig} to Dijkestra Guarded Commands are:

\[\AMDu\defs \begin{array}{l} \pchkdig \end{array} \]
\[\AMDma \defs\begin{array}{l} \pchkdigma \end{array} \]
\[\AMDmb \defs\begin{array}{l} \pchkdigmb \end{array} \]
\end{EExample}

\subsection{Dijkestra Weakest Precondition Predicate Transformer}
\label{sswp}
Dijkestra introduced the Weakest Precondition Predicate Transformer as a tool  for comparing the semantics of programs\cite{Dijkestra}.
The intuition behind this function is, if the weakest preconditions needed for executing the program $\PRg$ and terminating in  the state which is described by any post-conditions $\Gpost$ can be calculated ($\wpre{\PRg,\Gpost}$), then the semantic of $\PRg$ is understood. As a result, if $\wpre{\PRo,\Gpost} \iff \wpre{\PRt,\Gpost}$ for any $\Gpost$, then $\PRo$ is equivalent to $\PRt$.

In the rest of this subsection  the definitions of weakest precondition function ($\wpre{\PRg,\Gpost}$) when $\PRg$ is one of primitive programming statements are given.

\subsubsection{Weakest Preconditions of Primitive Programming Statements}

\renewcommand{\AMD}{Min}
\renewcommand{\AMDu}{\undertest{\AMD}}

\begin{Definition} $\daswp$
\label{daswp}

where $\Gpost\rN{w}{E}$ means substituting all free $w$ in $\Gpost$ with $E$.
\end{Definition}
Since $w$ will contain the value of $E$ after execution, then $\Gpost$ will be true after execution if and only if $\Gpost$, with the value of $w$ replaced by $E$, is true before execution.

\begin{Definition} $\dskwp$
\label{dskwp}
\end{Definition}
Execution of $\skipst$ does nothing, so, it is the identity transformation.

\begin{Definition} $\dscwp$
\label{dscwp}
\end{Definition}
Sequential composition is one way of composing larger program segments from smaller segments. It is executed by first executing $\PRo$ and then executing $\PRt$.

\begin{Definition}
 $\dnlwp$
\label{dnlwp}
\end{Definition}
The $\nullst$ is another name for $\skipst$ to be precise about the part of the program in which nothing is written. For example the part of the program before the first statement of the program is denoted by $\nullst$.

\begin{Definition}
 $\drtwp$
\label{drtwp}

where $\Gpn$ is the name of the program from which the execution is returned.
\end{Definition}
The $\returnst$  is treated as an assignment to the program name. 

\begin{Definition}
 $\dabwp$
\label{dabwp}
\end{Definition}

The statement $\abortst$ should never be executed, because it can only be executed in a state satisfying $\falsest$ and no state satisfies $\falsest$.

\begin{Definition}
$\difwp$
\label{difwp}
\end{Definition}
The first conjunct indicates that the guards must be well defined, and at least one guard is true. The second conjunct indicates  that execution of each program $\PRi$ with a true guard $\Gi$ terminates with $\Gpost$ true.

\begin{EExample} Consider the $\AMDu$ program given in Example \ref{exampmingc}.
\begin{argue}
\wpre{\AMDu, \Gpost} = \\
 \wpre{\pminab,\Gpost} =  & \Argdef{dscwp}\\
 \wpre{m:= a, \wpre{ \ifst b < a \rightarrow m := b \elsest b \geq a \rightarrow \skipst \fist, \wpre{\returnst(m), \Gpost}}}= & \Argdef{drtwp}\\
 \wpre{m:= a, \wpre{ \ifst b < a \rightarrow m := b \elsest b \geq a \rightarrow \skipst \fist, \Gpost\rN{\AMD}{m}}}= & \Argdef{difwp}\\
  \wpre{m:= a, (b<a \lor b \geq a) \land (b < a \implies \wpre{m := b, \Gpost\rN{\AMD}{m}}) \land (b \geq a \implies \wpre{\skipst, \Gpost\rN{\AMD}{m}})}= \\
  & \Arglaw{\lAonA},\Argdefs{dskwp}{daswp}\\
    \wpre{m:= a, (b < a \implies \Gpost\rN{\AMD}{m}\rN{m}{b}) \land (b \geq a \implies \Gpost\rN{\AMD}{m})}= & \Argdef{daswp}\\
     ((b < a \implies \Gpost\rN{\AMD}{m}\rN{m}{b}) \land (b \geq a \implies \Gpost\rN{\AMD}{m}))\rN{m}{a}= & \\
     (b < a \implies \Gpost\rN{\AMD}{m}\rN{m}{b})\rN{m}{a} \land (b \geq a \implies \Gpost\rN{\AMD}{m})\rN{m}{a}= & \\
     (b < a \implies \Gpost\rN{\AMD}{b}) \land (b \geq a \implies \Gpost\rN{\AMD}{a}) & \\     
\end{argue}
\end{EExample}

 \renewcommand{\AMD}{Chkdig}
\renewcommand{\AMDu}{\undertest{\AMD}}

\subsubsection{Weakest Preconditions of $\DO$ Statement}

$\DO$ is actually not a primitive construct, however, for simplicity, it is  treated  primitive. 
\begin{displaymath}
\DO \defs \begin{array}{l} \ydostv \end{array}
\end{displaymath}
A similar alternative statement can be defined with the same guarded command and $\Gdo$ which is the disjunction of the guards.
\begin{displaymath}
\IF \defs \begin{array}{l} \yifstv \end{array}
\end{displaymath}
It is obvious that the following iterative command is equivalent to $\DO$.
\begin{displaymath}
\DO \defs \dost \Gdo \rightarrow \left\{
\begin{array}{l} 
\yifstv 
\end{array}\right\}
\odst \\
\DO \defs \dost \Gdo \rightarrow \IF \odst
\end{displaymath}
This means, if all the guards are false, then $\Gdo$ is false and execution terminates; otherwise, the corresponding alternative command $\IF$ is executed and the process is repeated. One iteration of a loop, therefore, is equivalent to finding $\Gdo$ true and executing $\IF$. To formalize the weakest precondition of a loop, some terms must be defined first. 
\begin{Definition}
\label{ddowpa}
\begin{displaymath}
  \ddowpDa
\end{displaymath}
\end{Definition}

\begin{Definition}
\label{ddowpba}
$~$\\
\vspace{-.3in}
\ddowpDba
\end{Definition}

\begin{Definition}
\label{ddowpc}
\ddowpDc\\
\ddowpDce
\end{Definition}
\LWDC\, is based on the theorem which is given alongside its proof in Gries's book\cite{Gries},  and B-Method book\cite{Bbook} . The weakest precondition of a loop is defined by Definitions \ref{ddowpbb} and \ref{ddowp}.
\begin{Definition}
\label{ddowpbb}
$~$\\
\vspace{-.3in}
\ddowpDbb
\end{Definition}

\begin{Definition}
\label{ddowp}
$~$\\
 $\ddowpDdh$\\
\end{Definition}

In fact, first, the predicates given in the Loop Well-Definedness Conditions in Definition \ref{ddowpc} must be checked to understand that  $\Ido$, $\Vdo$ and $\Gdo$ are properly defined and the iterations of loop are not infinite.  Then the weakest precondition of loop can be calculated based on Definition \ref{ddowp}.

\newcommand{\ChkdigSTu}{\left\{\begin{array}{l} \pchkdigstuan \end{array}\right\}}
\newcommand{\ChkdigSTuif}{\left\{\begin{array}{l} \pchkdigstuifan \end{array}\right\}}
\newcommand{\ChkdigPRjb}{\pchkdigprjba}
\newcommand{\ChkdigPRjSTu}{\pchkdigprjstua}
\newcommand{\ChkdigPRja}{\pchkdigprjaa}
\newcommand{\ChkdigPRj}{\left\{\begin{array}{l} \ChkdigPRjb\\ \ChkdigPRjSTu \end{array}\right\}}
\newcommand{\ChkdigGj}{\pchkdiggja }
\newcommand{\ChkdigGdo}{\pchkdiggja}
\newcommand{\ChkdigGdoj}{\pchkdiggja}
\newcommand{\ChkdigIdo}{(0 \leq r \leq a) \land (d=0 \lor d=\lastd{r}) }
\newcommand{\ChkdigVdo}{r}
\newcommand{\ChkdigVardo}{t,r,d}

\begin{EExample} Consider the loop statement of $\AMDu$ program given in Example \ref{exampchkdiggc}.
\[\ChkdigSTu\]
To calculate the weakest precondition of this loop statement we need to clarify the following items.
\begin{Definition}
\label{dchkdigdef}
\begin{syntax}
\Gi  &:&   \ChkdigGj\\
\Gdo   &:& \ChkdigGdo\\
\PRi     &:& \ChkdigPRj\\
\Ido   &:& \ChkdigIdo\\
\Vdo   &:& \ChkdigVdo\\
\Vardo   &:& \ChkdigVardo\\
\end{syntax}
\end{Definition}
First, we must check the \LWDC\, conditions given in Definitions \ref{ddowpba} and \ref{ddowpc}.
\begin{argue}
\LWDC  & \Argdef{dchkdigdef}\\
\forall \ChkdigVardo @  (\ChkdigIdo \land \ChkdigGdoj \implies\\
\t3  \wpre{\ChkdigPRj,\ChkdigIdo}) \land  & \Argdef{dscwp}\\
\t2  (\ChkdigIdo \land \ChkdigGdoj \implies\\
\t3   \wpre{\Varfreshdo:=\ChkdigVdo \compose \ChkdigPRj, \ChkdigVdo < \Varfreshdo}))  \land & \Argdef{dscwp}\\
 \t2 (\ChkdigIdo \land \ChkdigGdo \implies \ChkdigVdo >0)=\\
\forall \ChkdigVardo @  (\ChkdigIdo \land \ChkdigGdoj \implies\\
\t3  \wpre{ t := r, \wpre{ r := \floor{\frac{t}{10}} , \wpre{ d := t-r*10,\ChkdigIdo}}}) \land  \\
& \Argdef{daswp}\\
\t2  (\ChkdigIdo \land \ChkdigGdoj \implies\\
\t3   \wpre{\Varfreshdo:=\ChkdigVdo, \wpre{  t := r, \wpre{ r := \floor{\frac{t}{10}}, \wpre{ d := t-r*10, \ChkdigVdo < \Varfreshdo}}}}))  \land & \Argdef{daswp}\\
\t2 (\ChkdigIdo \land \ChkdigGdo \implies \ChkdigVdo >0)=\\
\forall \ChkdigVardo @  (\ChkdigIdo \land \ChkdigGdoj \implies\\
\t3 \ChkdigIdo\rN{d}{ t-r*10}\rN{ r}{\floor{\frac{t}{10}}}\rN{ t}{r}) \land  \\
\t2  (\ChkdigIdo \land \ChkdigGdoj \implies\\
\t3  (\ChkdigVdo < \Varfreshdo)\rN{ d}{ t-r*10}\rN{ r}{ \floor{\frac{t}{10}}}\rN{t}{r}\rN{\Varfreshdo}{\ChkdigVdo}))  \land \\
 \t2 (\ChkdigIdo \land \ChkdigGdo \implies \ChkdigVdo >0)=\\
\forall \ChkdigVardo @  (\ChkdigIdo \land \ChkdigGdoj \implies\\
\t3 (0 \leq \floor{\frac{r}{10}} \leq a) \land (r-\floor{\frac{r}{10}}*10=0 \lor r-\floor{\frac{r}{10}}*10=\lastd{r})) \land  \\
\t2  (\ChkdigIdo \land \ChkdigGdoj \implies\\
\t3  (\floor{\frac{r}{10}} < r)))  \land \\
 \t2 (\ChkdigIdo \land \ChkdigGdo \implies \ChkdigVdo >0)=\\
 \truest\\
\end{argue}
Now, we calculate the weakest precondition of the loop statement:
\begin{argue}
 \wpre{\ChkdigSTu,\Gpost} =  & \Argdef{ddowp},\Argcalc{k=0}\\
 \t2  \DHp{0}{\Gpost}= (\Gpost \land \lnot \ChkdigGdo) \\
\end{argue}
\begin{argue}
 \wpre{\ChkdigSTu,\Gpost} =  & \Argdef{ddowp},\Argcalc{k=1}\\
 \t1 \DHp{1}{\Gpost} = & \Argdef{ddowp}\\
 \t1  (\wpre{\ChkdigSTuif, \DHp{0}{\Gpost}})=  & \Argdef{difwp}\\
\t1 (\pchkdiggja \lor \lnot \pchkdiggja) \land &\Arglaw{\lAonA}\\
\t2  ~~  \pchkdiggja \implies  \wpre{t := r\compose r := \floor{ \frac{t}{10}}  \compose  d := t-r*10, \DHp{0}{\Gpost} } \land &  \Argdef{dscwp}\\
\t2   \lnot \pchkdiggja \implies \wpre{ \skipst, \DHp{0}{\Gpost}}= & \Argdef{dskwp}\\
\t1 \truest \land\\
\t1    \pchkdiggja \implies  \wpre{t := r, \wpre{ r := \floor{ \frac{t}{10}},   \wpre{ d := t-r*10,  \DHp{0}{\Gpost} }}} \land & \Argdef{daswp}\\
\t1   \lnot \pchkdiggja \implies \DHp{0}{\Gpost} =\\
\t1    \pchkdiggja \implies \DHp{0}{\Gpost} \rN{ d}{ t-r*10}\rN{ r}{ \floor{ \frac{t}{10}} }\rN{t}{ r } \land \\
\t1   \lnot \pchkdiggja \implies \DHp{0}{\Gpost}= & \Argcalc{\DHp{0}{\Gpost} = (\Gpost \land \lnot \ChkdigGdo)}\\
\t1    \pchkdiggja \implies  (\Gpost \land \lnot( r > 0 \land d < b)) \rN{ d}{ t-r*10}\rN{ r}{ \floor{ \frac{t}{10}} }\rN{t}{ r } \land \\
\t1   \lnot \pchkdiggja \implies  (\Gpost \land \lnot ( r > 0 \land d < b))=\\
\t1    \pchkdiggja \implies  (\Gpost \land \lnot ( r > 0 \land (r-\floor{ \frac{r}{10}}*10) < b))  \land \\
\t1   \lnot \pchkdiggja \implies  (\Gpost \land \lnot ( r > 0 \land d < b))=\\
\end{argue}
\end{EExample}

%% file: zp0200.tex
\setcounter{chap}{2}
\section{Literature Survey}
To our knowledge there is not any formal approach to calculate Reachability, Infection and Propagation conditions in the literature.  What are discussed in various literature (e.g. \cite{BuddAngluin1982}, \cite{LarryMorell1984}, \cite{RichardsonThompson1988}, \cite{Morell1988}, \cite{DeMilliOffutt1991}, and \cite{VoasMiller1992}) are defining these concepts  with various names (e.g. creation instead of infection  in \cite{LarryMorell1984}, necessity instead of infection and sufficiency instead of propagation in \cite{DeMilliOffutt1991}, execution  instead of reachability in \cite{VoasMiller1992}) and using them to generate test cases automatically (e.g. in \cite{DeMilliOffutt1991}) or proving the correctness of programs(e.g. in \cite{BuddAngluin1982}). The work which is very closed to our work is \cite{DeMilliOffutt1991}, in which some infection constraint templates for the effect of each one of mutation operators are given.

%% file: zp0300.tex
\setcounter{chap}{3}
\section{The proposed method}
\label{propsec}

In this section the  proposed method is discussed. First, the understandings of the Reachability, Infection and Propagation conditions must be clarified.
To calculate reachability of a mutated statement, the condition generated by the execution of each one of primitive statements which are before the mutated statement must be calculated. This concept is discussed in Subsection \ref{ssrcgf}. To calculate the Infection condition, the semantic differences between  the execution of the original statement and the execution of mutated statement must be investigated. This is done with the help of the weakest precondition function defined in Subsection \ref{sswp}. Calculating the Propagation condition is a troublesome part of this study. It is obvious that calculating the conditions needed to reach to the end of the program under test, is not that much helpful. What is has to be known again is the semantic differences between two execution paths. One path is, starting the execution from  the mutated statement  to the end of program, and the other path is, starting the execution from the original statement to the end of program. This is done, again, with the help of the weakest precondition function. In Subsection \ref{sscpm}, first, two types of modification of the program in mutation testing are distinguished, and the original and mutated program templates are introduced. An algorithmic incremental  method is introduced to calculate Reachability, Infection and Propagation conditions based on these templates in Subsection \ref{ssripcc}.  

\subsection{Reachability Condition Generator Function}
\label{ssrcgf}
 The reachability condition generator function, namely $\reachc{prog}$, is introduced for calculating the reachability condition generated by $prog$.   
 The definitions of reachability condition function applying on primitive programming statements are as follow:

If the mutated statement is located after an assignment, a $\skipst$, or a $\nullst$ statement, no condition is needed to reach to that statement. So the reachability condition generated by these three statements are $\truest$.
\begin{Definition} $\dasrc$
\label{dasrc}
\end{Definition}

\begin{Definition} $\dskrc$
\label{dskrc}
\end{Definition}

\begin{Definition}
 $\dnlrc$
\label{dnlrc}
\end{Definition}

If before reaching to a mutated statement a $\returnst$ statement or an $\abortst$ statement is executed there is no way to reach to that statement. So the reachability condition generated by these two statements are $\falsest$.
\begin{Definition}
 $\drtrc$
\label{drtrc}
\end{Definition}

\begin{Definition}
 $\dabrc$
\label{dabrc}
\end{Definition}

The reachability condition generated by two sequentially composed statements is the conjunction of two reachability conditions generated by each one of the statements.
\begin{Definition} $\dscrc$
\label{dscrc}
\end{Definition}
An alternative statement generates the reachability condition as the disjunction of its true guards which are joined by conjunction to  the reachability condition generated by the corresponding guarded commands.
\begin{Definition}
$~$\\
 $\difrc$
\label{difrc}
\end{Definition}

 The same as what is defined to formulate the weakest precondition of a loop, to formalize the reachability condition of a loop, in addition to what is defined in Definitions \ref{ddowpa} and \ref{ddowpc},  some terms must be defined: 
\begin{Definition}
\label{ddorcb}
$~$\\
\vspace{-.3in}
\ddorcMb
\end{Definition}

\begin{Definition}
$~$\\
 $\ddorcMd$
\label{ddorc}
\end{Definition}
If the $\LWDC$ condition (Definition \ref{ddowpc}) for the loop is $\falsest$,  then the reachability condition generated by the loop is $\falsest$. 

Discussion: Assume the mutated statement is located after a loop statement. To calculate the reachability condition of the mutated statement, the minimum value for $k$, that is the number of iterations of the loop, must be find. This is the number of iterations in which the loop terminates, and we reach to the mutated statement. This means, if this is possible to not enter the loop, that is $k=0$ and means  executing the loop with zero iteration, the reachability condition generated by the loop is $\MRp{0}$ which is $\lnot \Gdo$. But, the problem is when $\Gdo$ cannot be $\falsest$ at the beginning and the loop body has to be executed one or more than once. In this case, based on the body of the loop three cases are distinguished:
\begin{enumerate}
\item The body of the loop is the composition of assignment, $\skipst$ or $\nullst$ statements. None of these primitive statements generate any conditions. As a result, there is no difference between $k=0$ and $k\geq 1$.
\item If the body of the loop is the composition of statements mentioned in case 1 and  $\returnst$ or $\abortst$, the condition which is generated is $\falsest$ and it means there is no way to the mutated statement from this branch. 
\item If the body of the loop consists of an alternation statement or a loop statement, it is needed to consider their guards as the condition generated by the body of the loop. So if the loop terminates after one iteration, that is $k=1$, then the reachability condition generated by the loop is $\MRp{1}$. But what about for $k >1$? What can be suggested is,  if a mutated statement which is located after a loop statement cannot be reached with zero or at most one iteration of the loop, it  can be assumed that the reachability condition generated by the loop statement is $\falsest$ and as a result, this mutated statement cannot be reached, at least easily,  via this branch of the program under test. 
\end{enumerate}
 \renewcommand{\AMD}{Chkdig}
\renewcommand{\AMDu}{\undertest{\AMD}}
\renewcommand{\AMDma}{\mutateda{\AMD}}
\renewcommand{\AMDmb}{\mutatedb{\AMD}}
\begin{EExample}
Consider the second mutant $\AMDmb$ of the $\AMD$ program of Example \ref{exampchkdiggc}. To calculate the reachability condition of the mutated statement which located after a loop, it is needed to calculate the reachability condition generated by the loop. That is:
\begin{argue}
\reachc{\pchkdigstuan} = & \Argdef{ddorc}\\
\t1 \exists k :\nat @ \MRp{k}= & \Argcalc{k=0}\\
\t1 \MRp{0} =  \lnot \Gdo=\lnot (r > 0 \land d < b)=\\
\t1 r \leq 0 \lor d \geq b = & \Argcalc{k=0 \implies r=0}
\end{argue}
Which means, if the input ($a$) is zero the execution does not enter the loop and  the mutated statement is reached.
\end{EExample}

\subsection{Categorizing Program Modifications}
\label{sscpm}
Two types of modifications are distinguished, namely,  the {\it statement modification} and the {\it guarded command modification}. Definition \ref{putdef} specifies a template for an under test  module ($\MDsu$)\footnote{See Table \ref{natable} for naming and abbreviation standards}   and Definition \ref{mutdef} specifies a template for its mutated version ($\MDsm$) when  the statement modification is applied.

\begin{Definition}
 $\MDsu \defs \PRb  \compose \STu  \compose \PRa$
\label{putdef}
\end{Definition}
Each under test module ($\MDsu$) is a composition of three parts. The part of the module which is before the statement which must be modified ($\PRb$), the statement which must be modified($\STu$), and  the part of the module which is after the statement which must be modified ($\PRa$). It is obvious that $\PRb$ and $\PRa$ can be $\nullst$.

\begin{Definition} $\MDsm \defs \PRb  \compose \STm  \compose \PRa$ \\
\hspace*{.2in} while $\STm$ is the modified version of  $\STu$ of Definition \ref{putdef} which is modified by one of the mutation operators.
\label{mutdef}
\end{Definition}
When the  module $\MDsu$ is modified ($\MDsm$), it is the statement ($\STu$) part which is modified ($\STm$) and the other parts remain unchanged.

The  statement modifications which are done by the help of mutation operators can be categorized as shown in Table \ref{msttable}. If one of the guards of  alternative or iterative statements is modified it is assumed as the statement modification.  
\begin{table}[h!]
\caption{Statement modifications}
\centering
$
\begin{array}{|l|l|l|}
\hline
\mbox{Statement Under Test}(\STu) & \mbox{Mutated Statement}(\STm) & \mbox{Comments}\\
\hline
\yasst  &  w := \mutated{E} &  \\
\hline
\yasst  &  \mutated{w} := E & \\
\hline             
\ymast   & w_j:= \mutated{E_{j}} &  \\
\hline
  \ymast   & \mutated{w_{j}}:=  E_{j}&  \\
\hline  

 \begin{array}{l}
  \yifstv             
 \end{array}
  &   \begin{array}{l}
                        \ifst \Go \rightarrow \PRo\\
                        \vdots\\
                        \elsest \mutated{\Gj} \rightarrow \PRj\\
                        \vdots\\
                         \elsest \Gn \rightarrow \PRn\\
                         \fist\\                        
                \end{array}
                  &  \\

\hline  
\begin{array}{l}
\ydostv
\end{array} &   \begin{array}{l}
\dost \Go \rightarrow \PRo\\
\vdots\\
\elsest \mutated{\Gj} \rightarrow \PRj\\
\vdots\\
\elsest \Gn \rightarrow \PRn\\
\odst
\end{array} & \\
\hline
\end{array}
$
\label{msttable}
\end{table}
\newcommand{\yifdosthnu}{\ifst|\dost \\
\t1 \elsest_{i\neq j} . \Gi \rightarrow \PRi \\
\t1 \elsest \Gj \rightarrow \PRju \\
\fist|\odst}
\newcommand{\yifdosthnm}{\ifst|\dost \\
\t1 \elsest_{i\neq j} . \Gi \rightarrow \PRi \\
\t1 \elsest \Gj \rightarrow \PRjm \\
\fist | \dost}
\newcommand{\yifsthnu}{\ifst \\
\t1 \elsest_{i\neq j} . \Gi \rightarrow \PRi \\
\t1 \elsest \Gj \rightarrow \PRju \\
\fist}
\newcommand{\yifsthnm}{\ifst \\
\t1 \elsest_{i\neq j} . \Gi \rightarrow \PRi \\
\t1 \elsest \Gj \rightarrow \PRjm \\
\fist}
\newcommand{\ydosthnu}{\dost \\
\t1 \elsest_{i\neq j} . \Gi \rightarrow \PRi \\
\t1 \elsest \Gj \rightarrow \PRju \\
\odst}
\newcommand{\ydosthnm}{\dost \\
\t1 (\elsest_{i\neq j} . \Gi \rightarrow \PRi \\
\t1 \elsest \Gj \rightarrow \PRjm) \\
\odst}

Definition \ref{putdefn} specifies a template for an under test  module ($\MDgu$)  and Definition \ref{mutdefn} specifies a template for its mutated version ($\MDgm$) when  the guarded command modification is applied. In these definitions,  $\ifst|\dost$  means that the statement can be an alternative statement or an iterative statement.
\begin{Definition}
 $\MDgu \defs \PRb  \compose
\left\{\begin{array}{l}
 \yifdosthnu
  \end{array}\right\} \compose  \PRa
 $
\label{putdefn}
\end{Definition}
Each under test module ($\MDgu$) is a composition of three parts. The part of the module which is before $\ifst|\dost$ statement ($\PRb$), the $\ifst|\dost$ statement which its $j^{th}$ guarded command must be modified ($\PRjb \compose  \STju \compose  \PRja $), and  the part of the module which is after the  $\ifst|\dost$ statement ($\PRa$). As it can be seen, the $j^{th}$ guarded command of the middle part again is  composition of three parts, namely,  $\PRjb$, $\STju$, and  $\PRja$.
\begin{Definition} 
$\MDgm \defs \PRb  \compose
\left\{\begin{array}{l}
 \yifdosthnm
  \end{array}\right\} \compose  \PRa
 $
 
\hspace*{.2in} while $\STjm$ is the modified version of  $\STju$ of Definition \ref{putdefn} which is modified by one of the mutation operators.
\label{mutdefn}
\end{Definition}
When the  module $\MDgu$ is modified ($\MDgm$), it is the statement ($\STju$) part of its $j^{th}$ guarded command which is modified ($\STjm$) and the other parts remain unchanged.

In the  guarded command  modifications, the modification which is applied on $\STju$  by the help of mutation operators  can be categorized again as the statement modifications or the guarded  command modifications. In the case of statement modifications, it follows the categorize are shown in Table \ref{msttable}. In the case of guarded command modifications it follows the Definitions \ref{putdefn} and \ref{mutdefn}.

\subsection{Reachability, Infection, and Propagation Conditions Calculations}
\label{ssripcc}
The Reachability, Infection and Propagation conditions when the statement modifications are applied, based on the Weakest Precondition  and Reachability Condition definitions given in Section 1,  and Reachability Condition definitions given in Subsection \ref{ssrcgf}, are given in Definitions \ref{rchdef}, \ref{infdef} and \ref{ppgdef}, respectively.
 
\newcommand{\Reachabilitys}{\stmod{Reachability}}
\newcommand{\Infections}{\stmod{Infection}}
\newcommand{\Propagations}{\stmod{Propagation}}
\newcommand{\FTSrips}{\stmod{Full\_Test\_Specification}}
\newcommand{\Reachabilityg}{\gdmod{Reachability}}
\newcommand{\Infectiong}{\gdmod{Infection}}
\newcommand{\Propagationgif}{\gdmod{Propagation^{if}}}
\newcommand{\Propagationgdo}{\gdmod{Propagation^{do}}}
\newcommand{\FTSripgif}{\gdmod{Full\_Test\_Specification^{if}}}
\newcommand{\FTSripgdo}{\gdmod{Full\_Test\_Specification^{do}}}
\newcommand{\Reachability}{\reachc{\PRb}}
\begin{Definition} $\Reachabilitys(\STu$):\\
\label{rchdef}
\hspace*{.5in} {\Large $\Reachability $}
\end{Definition}

Based on the template given in Definition \ref{putdef}, the reachability condition for $\STu$ is the reachability condition generated by the statements of the $\PRb$.

\newcommand{\Infection}{\wpre{\STu,\Gpost} \neq \wpre{\STm,\Gpost}}
\begin{Definition} $\Infections(\STu,\STm$):\\
\hspace*{.5in} {\Large $\Infection$}
\label{infdef}
\end{Definition}

The infection condition when $\STu$ is modified to $\STm$ is that the conditions on which the  execution of $\STu$ differs from the execution of $\STm$.

\newcommand{\Propagation}{
\wpre{\STu \compose \PRa,\Gpost}  \neq \wpre{ \STm \compose \PRa,\Gpost}
}
\begin{Definition} $\Propagations(\STu,\STm$) : \\
\hspace*{.5in} {\Large $\Propagation$}
 \label{ppgdef}
\end{Definition}

The propagation condition when $\STu$ is modified to $\STm$ is that the conditions on which the  execution of $\STu \compose \PRa$ differs from the execution of $\STm \compose \PRa$.
\newcommand{\FTSrip}{\Reachabilitys(\STu) \land \Infections(\STu,\STm) \land \Propagations(\STu,\STm)}
\begin{Definition} $\FTSrips(\STu, \STm$):\\
\hspace*{.5in} {\Large $\FTSrip$ }
\label{ftsdefrip}
\end{Definition}
 
 The full test specification when $\STu$ is modified to $\STm$ is  the conjunction of the reachability, infection and propagation conditions.
 
When the guarded command modifications are applied the reachability, infection and propagation  conditions are given in Definitions \ref{rchdefn}, \ref{infdefn}, \ref{ppgdefnif}, and \ref{ppgdefndo}.

\newcommand{\Reachabilityn}{\reachc{\PRb} \land  \Gj \land  \reachc{\PRjb}}
\begin{Definition} $\Reachabilityg(\STju$):\\
\label{rchdefn}
\hspace*{.5in} {\Large $\Reachabilityn $} \\ 
\end{Definition} 

Based on the template given in Definition \ref{putdefn}, the reachability condition for $\STju$ is the conjunction of the reachability condition generated by the statements of the $\PRb$ and $j^{th}$ guard ($G_j$)  and the reachability condition generated by the statements of the $\PRjb$. if  $\STju$ is again a $\ifst|\dost$ statement then $\Gj$ is the conjunction of all guards, and $\PRjb$  is the composition of all programs ($\PRjb$s)  from the outer  statement to the inner statement.

\newcommand{\Infectionn}{\wpre{\STju,\Gpost} \neq \wpre{\STjm,\Gpost}}
\begin{Definition} $\Infectiong(\STju,\STjm$):\\
\hspace*{.5in} {\Large $\Infectionn$}
\label{infdefn}
\end{Definition}

The infection condition when $\STju$ is modified to $\STjm$ is that the  conditions on which the execution of $\STju$ differs from the execution of $\STjm$.

\newcommand{\Propagationnif}{
\wpre{\STju \compose \PRja \compose \PRa,\Gpost}  \neq \wpre{ \STjm \compose \PRja \compose \PRa,\Gpost}
}
\begin{Definition} $\Propagationgif(\STju, \STjm$) : \\
\hspace*{.5in} {\Large $\Propagationnif$}
 \label{ppgdefnif}
\end{Definition}

The propagation condition when $\STu$ is an alternate statement and $\STju$ is modified to $\STjm$ is that the conditions on which the execution of $\STju \compose \PRja \compose  \PRa$ differs from the execution of $\STm \compose \PRja \compose \PRa$. For simplicity, one level of nesting is assumed, but it is obvious that if  $\STju$ is again a $\ifst$ statement then  $\PRja$  is the composition of all programs ($\PRja$s)  from the inner  statement to the outter statement.

\newcommand{\Propagationndo}{
\wpre{\STu \compose  \PRa,\Gpost}  \neq \wpre{ \STm  \compose \PRa,\Gpost}
}
\begin{Definition} $\Propagationgdo(\STju, \STjm$) : \\
\hspace*{.5in} {\Large $\Propagationndo$}
 \label{ppgdefndo}
\end{Definition}

The propagation condition when $\STu$ is an iterative statement and $\STju$ is modified to $\STjm$ is that the conditions on which the execution of $\STu \compose  \PRa$ differs from the execution of $\STm \compose  \PRa$. 

\newcommand{\FTSripnif}{\Reachabilityg(\STju) \land \Infectiong(\STju,\STjm) \land \Propagationgif(\STju,\STjm)}
\newcommand{\FTSripndo}{ \Reachabilityg(\STju) \land \Infectiong(\STju,\STjm) \land \Propagationgdo(\STju,\STjm)}

\begin{Definition} $\FTSripgif(\STju, \STjm$):\\
\hspace*{.5in} {\Large $\FTSripnif$ }
\label{ftsdefripnif}
\end{Definition}

\begin{Definition} $\FTSripgdo(\STju, \STjm$):\\
\hspace*{.5in} {\Large $\FTSripndo$ }
\label{ftsdefripndo}
\end{Definition}
  The full test specification when $\STju$ is modified to $\STjm$ is  the conjunction of the reachability, infection and propagation conditions.

In the next section, these definitions are applied on some case studies to show the applicability of the method.

%% file: zp0401.tex
\setcounter{chap}{4}
\section{Case studies}
\label{scase}
In this section, the proposed method is applied to four cases, namely, $Min$, $isEven$, $Chkdig$, and $Search$.  The programs $Min$, $isEven$, and  $Chkdig$ are used as running examples throughout the previous sections. 
In order that this section to be self-contained, source codes and mutants of these three programs are repeated. The program $Search$ is discussed totally in the section. 
\subsection{Min Program: Killing Mutant Strongly and Weakly}
\subsubsection{Definitions}
The original code in which the mutated statements are marked by prime, discussed in Example \ref{exampmin},  is as follow:

\pminabvb

Abbreviating $minVal$ by $m$, $A$ by $a$ and $B$ by $b$,
 \, the translation of this program and its mutated version for Mutant 1 to Dijkestra Guarded Commands are (Example \ref{exampmingc}):
 \renewcommand{\AMD}{Min}
\renewcommand{\AMDu}{\undertest{\AMD}}
\renewcommand{\AMDma}{\mutateda{\AMD}}
\[\AMDu\defs \pminab \]
\[\AMDma \defs\pminabma \]

\newcommand{\MinPRb}{\nullst}
\newcommand{\MinSTu}{m:=a}
\newcommand{\MinPRa}{ \ifst b < a \rightarrow m := b \elsest b \geq a \rightarrow \skipst \fist\compose \returnst(m)}
\newcommand{\MinSTm}{\mutatedpart{m:=b}}

 The underlined statement is the mutated statement. Since this is a statement modification, to calculate the RIP conditions, the above programs and what are given in Definitions \ref{putdef} and \ref{mutdef} must be  compared. So:\\
\begin{Definition} 
\label{minadef}
\begin{syntax}
\PRb & : &  \MinPRb  \\
\STu & : & \MinSTu   \\
\PRa & : & \MinPRa \\
\STm & : &  \MinSTm
\end{syntax}
\end{Definition}

\subsubsection{RIP Calculations}

\begin{argue}
\underline{{\bf Reachability}}: &\Argdef{rchdef}\\
\Reachabilitys(\STu)= \Reachability &  \Argdef{minadef}\\
\Reachabilitys(\MinSTu)=  \\
\t1 \reachc{\MinPRb} = &\Argdef{dnlrc}\\
\t1 \truest
\end{argue}

\begin{argue}
\underline{{\bf Infection}}: &  \Argdef{infdef}\\
\Infections(\STu,\STm) = \Infection &  \Argdef{minadef}\\
\Infections(\MinSTu,\MinSTm) = \\
\t1  \wpre{\MinSTu,\Gpost} \neq \wpre{\MinSTm,\Gpost}  & \Argdef{daswp}\\
\t1  \Gpost\rN{m}{a} \neq \Gpost\rN{m}{b}\\
\t1  a \neq b
 \end{argue}
 So if $a \neq b$ the $\AMDma$ will be killed weakly.

    \begin{argue}
\underline{{\bf Propagation}}: & \Argdef{ppgdef}\\   
 \Propagations(\STu,\STm)= \Propagation & \Argdef{dscwp}\\
 \Propagations(\STu,\STm)= \wpre{\STu, \wpre{\PRa,\Gpost}} \neq \wpre{\STm, \wpre{\PRa,\Gpost}}=  &  \\
\end{argue}
\begin{argue}
\wpre{\PRa,\Gpost}=  & \Argdef{minadef}\\
 \t1 \wpre{\MinPRa,\Gpost}= & \Argdef{dscwp}\\
\t1    \wpre{\ifst b < a \rightarrow m := b \elsest b \geq a \rightarrow \skipst \fist, \wpre{\returnst(m),\Gpost}} = &  \Argdef{drtwp}\\
\t1    \wpre{\ifst b < a \rightarrow m := b \elsest b \geq a \rightarrow \skipst \fist, \Gpost\rn{\AMDu}{m}} = &  \Argdef{difwp}\\
\t1   (b<a) \lor (b \geq a) \land   (b < a \implies \wpre{m := b, \Gpost\rn{\AMDu}{m}}) \land (b \geq a \implies\wpre{\skipst,\Gpost\rn{\AMDu}{m}}) = & \\
&   \Arglaw{\lAonA},\Argdefs{dskwp}{daswp}\\
\t1  \truest \land   (b < a \implies \Gpost\rn{\AMDu}{m}\rN{m}{b}) \land (b \geq a \implies \Gpost\rn{\AMDu}{m}) = & \\
\t1  (b < a \implies \Gpost\rn{\AMDu}{b}) \land (b \geq a \implies \Gpost\rn{\AMDu}{m})  & \\
\end{argue}
\begin{argue}
 \wpre{\STu, \wpre{\PRa,\Gpost}} = &  \Argcalc{\uparrow},\Argdef{minadef}\\ 
\t1   \wpre{\MinSTu,(b < a \implies \Gpost\rn{\AMDu}{b}) \land (b \geq a \implies \Gpost\rn{\AMDu}{m}) } = & \Argdef{daswp}\\
\t1 ((b < a \implies \Gpost\rn{\AMDu}{b}) \land (b \geq a \implies \Gpost\rn{\AMDu}{m}))\rN{m}{a}  = & \\
\t1 (b < a \implies \Gpost\rn{\AMDu}{b}) \land (b \geq a \implies \Gpost\rn{\AMDu}{a})  & 
\end{argue}
\begin{argue}
 \wpre{\STm, \wpre{\PRa,\Gpost}} = &  \Argcalc{\uparrow},\Argdef{minadef}\\ 
\t1   \wpre{\MinSTm,(b < a \implies \Gpost\rn{\AMDma}{b}) \land (b \geq a \implies \Gpost\rn{\AMDma}{m}) } = & \Argdef{daswp}\\
\t1 ((b < a \implies \Gpost\rn{\AMDma}{b}) \land (b \geq a \implies \Gpost\rn{\AMDma}{m}))\rN{m}{b}  = & \\
\t1 (b < a \implies \Gpost\rn{\AMDma}{b}) \land (b \geq a \implies \Gpost\rn{\AMDma}{b})  & 
\end{argue}

  \begin{argue}
  \Propagations(\MinSTu,\MinSTm)=\Propagation \\
\t1   (b < a \implies \Gpost\rn{\AMDu}{b}) \land (b \geq a \implies \Gpost\rn{\AMDu}{a})  \neq \\
       \t1 (b < a \implies \Gpost\rn{\AMDma}{b}) \land (b \geq a \implies \Gpost\rn{\AMDma}{b})
  \end{argue}
  This shows that if $b < a$ then the result of $\AMDma$ is the same as the result of $\AMDu$, but when $b \geq a $ then these two results are different. As a result, the propagation condition is $b \geq a$.

 \begin{argue}
 \underline{{\bf Full~ Test~ Specification}}: &  \Argdef{ftsdefrip}\\
 \FTSrips(\STu,\STm) = \\
 \FTSrip = & \Argcalc{\uparrow}\\
 true \land a \neq b \land  b \geq a =\\
 b > a  
 \end{argue}
For $b < a$ both programs have the same result. \\
For $ b \geq a$, the condition is $\Gpost\rN{\AMDu}{m}\rN{m}{a} \neq \Gpost\rN{\AMDma}{m}\rN{m}{b}$. This means $a \neq b$. As a result,  $b > a$. Therefore, if test cases are selected in which $b > a$ then    $\AMDma$ will be killed strongly. 

%% file: zp0402.tex
\subsection{Min Program: Original Program and the Mutant are Equivalent}
\subsubsection{Definitions}
The original code in which the mutated statements are marked by prime, discussed in Example \ref{exampmin}, is as follow:

\pminabvb

The translation of this program and its mutated version for Mutant 2  to Dijkestra Guarded Commands  with the same abbreviations as before are (Example \ref{exampmingc}):
\renewcommand{\AMDmb}{\mutatedb{\AMD}}
\[\AMDu\defs \pminab \]
\[\AMDmb\defs\pminabmb \]

\renewcommand{\MinPRb}{m:=a}
\renewcommand{\MinSTu}{\ifst b < a \rightarrow m := b \elsest b \geq a \rightarrow \skipst \fist\compose \returnst(m)}
\renewcommand{\MinPRa}{\nullst }
\renewcommand{\MinSTm}{\ifst \mutatedpart{b < m} \rightarrow m := b \elsest \mutatedpart{b \geq m} \rightarrow \skipst \fist\compose \returnst(m)}

The underlined condition is the mutated part of program. Since this is a statement modification,  the above programs and what are given in Definitions \ref{putdef} and \ref{mutdef} must be compared. So:\\
\begin{Definition} 
\label{minbdef}
\begin{syntax}
\PRb & : &  \MinPRb  \\
\STu & : & \MinSTu   \\
\PRa & : & \MinPRa \\
\STm & : &  \MinSTm
\end{syntax}
\end{Definition}

\subsubsection{RI Calculations}

\begin{argue}
 \underline{{\bf Reachability}}: &\Argdef{rchdef}\\
 \Reachabilitys(\STu)=  \Reachability  &  \Argdef{minbdef}\\
 \Reachabilitys(\MinSTu)=    &  \\
\t1 \reachc{\MinPRb} = & \Argdef{dasrc}\\
\t1 \truest
\end{argue}
 
\begin{argue}
 \underline{{\bf Infection}}:   & \Argdef{infdef}\\
\Infections(\STu,\STm)=  \Infection & \\
\wpre{\STu,\Gpost}= &  \Argdef{minbdef}\\
\t1  \wpre{\MinSTu,\Gpost} & \Argdefs{drtwp}{difwp}\\
\t1  (b <a \lor b \geq a) \land (b <a \implies \wpre{m:=b,\Gpost\rN{\AMDu}{m}}) \land (b \geq a \implies \wpre{\skipst,\Gpost\rN{\AMDu}{m}}) \\
  & \Arglaw{\lAonA}, \Argdefs{daswp}{dskwp}\\
\t1 \truest \land  (b <a \implies \Gpost\rN{\AMDu}{m}\rN{m}{b}) \land (b \geq a \implies \Gpost\rN{\AMDu}{m}) & \\
\t1  (b <a \implies \Gpost\rN{\AMDu}{m}\rN{m}{b}) \land (b \geq a \implies \Gpost\rN{\AMDu}{m}) & \\
\end{argue}
\begin{argue}
\wpre{\STm,\Gpost}=&  \Argdef{minbdef}\\
 \t1  \wpre{\MinSTm,\Gpost}=  & \Argdefs{drtwp}{difwp}\\
\t1 (b <m \lor b \geq m) \land (b <m \implies \wpre{m:=b,\Gpost\rN{\AMDmb}{m}}) \land (b \geq m \implies \wpre{\skipst,\Gpost\rN{\AMDmb}{m}})\\
  & \Arglaw{\lAonA}, \Argdefs{daswp}{dskwp}\\
\t1 \truest \land (b <m \implies \Gpost\rN{\AMDmb}{m}\rN{m}{b}) \land (b \geq m \implies \Gpost\rN{\AMDmb}{m})  & \\
\t1 (b <m \implies \Gpost\rN{\AMDmb}{m}\rN{m}{b}) \land (b \geq m \implies \Gpost\rN{\AMDmb}{m})  & \\
\end{argue}
\begin{argue}
\Infection\\
\t1 (b <a \implies \Gpost\rN{\AMDu}{m}\rN{m}{b}) \land (b \geq a \implies \Gpost\rN{\AMDu}{m}) \neq\\
\t1  (b <m \implies \Gpost\rN{\AMDmb}{m}\rN{m}{b}) \land (b \geq m \implies \Gpost\rN{\AMDmb}{m}) 
& \Arglaw{\MinPRb}\\
\t1 (b <a \implies \Gpost\rN{\AMDu}{m}\rN{m}{b}) \land (b \geq a \implies \Gpost\rN{\AMDu}{a}) \neq\\
\t1  (b <a \implies \Gpost\rN{\AMDmb}{m}\rN{m}{b}) \land (b \geq a \implies \Gpost\rN{\AMDmb}{a}) \\
\t1 \falsest
 \end{argue}
 So  the $\AMDmb$ is equivalent to $\AMDu$.

%% file: zp0403.tex
\subsection{isEven Program: Killing Mutant Strongly and Weakly}
\subsubsection{Definitions}
The original code in which the mutated statement is marked by prime, discussed in Example \ref{exampiseven}, is as follow:

\pisevenvb

Abbreviating $X$ by $a$ ,
 \, the translation of this program and its mutated version to Dijkestra Guarded Commands are (Example \ref{exampisevengc}):
\renewcommand{\AMD}{isEven}
\renewcommand{\AMDu}{\undertest{\AMD}}
\renewcommand{\AMDm}{\mutated{\AMD}}
\[\AMDu\defs \begin{array}{l} \piseven \end{array} \]
\[\AMDm \defs\begin{array}{l} \pisevenm \end{array} \]

\newcommand{\isEvenPRb}{\nullst}

\newcommand{\isEvenSTu}{\ifst a < 0 \rightarrow a := 0- a \elsest a \geq 0  \rightarrow \skipst \fist}
\newcommand{\isEvenPRjb}{\nullst}
\newcommand{\isEvenGj}{a <0 }
\newcommand{\isEvenPRjSTu}{a := 0- a}
\newcommand{\isEvenPRja}{\nullst}
\newcommand{\isEvenPRa}{ \ifst \floor{\frac{a}{2}} = \frac{a}{2} \rightarrow \returnst(\truest) \elsest \floor{\frac{a}{2}} \neq \frac{a}{2} \rightarrow \returnst(\falsest) \fist}
\newcommand{\isEvenSTm}{\ifst a < 0 \rightarrow \mutatedpart{a := 0} \elsest a \geq 0  \rightarrow \skipst \fist}
\newcommand{\isEvenPRjSTm}{a := 0}

  The underlined statement is the mutated statement. This is a guarded command modification. To calculate the RIP conditions,  the above programs and what are given in Definitions \ref{putdefn} and \ref{mutdefn} must be compared. So:\\
\begin{Definition} 
\label{isevendef}
\begin{syntax}
\PRb & : &  \isEvenPRb  \\
\STu & : & \isEvenSTu   \\
\PRjb &:& \isEvenPRjb\\
\STju & : & \isEvenPRjSTu   \\
\PRja &:& \isEvenPRja\\
\PRa & : & \isEvenPRa \\
\STm & : &  \isEvenSTm\\
\STjm & : &  \isEvenPRjSTm\\
\Gj  &:&   \isEvenGj\\
\end{syntax}
\end{Definition}

\subsubsection{RIP Calculations}

\begin{argue}
 \underline{{\bf Reachability}}: &\Argdef{rchdefn}\\
\Reachabilityg(\STju)=  \Reachabilityn   & \Argdef{isevendef}\\
\Reachabilityg(\isEvenPRjSTu)=   &\\
\t1 \reachc{\isEvenPRb} \land a<0 \land \reachc{\isEvenPRjb} = & \Argdef{dnlrc}\\
\t1 \truest \land  a < 0  \land \truest & \\
\t1 a < 0
\end{argue}

\begin{argue}
 \underline{{\bf Infection}}:  & \Argdef{infdefn}\\
\Infectiong(\STju,\STjm)= \Infectionn & \\
\wpre{\STju,\Gpost}=  & \Argdef{isevendef}\\
\t1  \wpre{\isEvenPRjSTu,\Gpost} &  \Argdef{daswp} \\
\t1  \Gpost\rN{a}{0-a}) \\
 \end{argue}
 \begin{argue}
  \wpre{\STjm,\Gpost}= & \Argdef{isevendef}\\
\t1 \wpre{\isEvenPRjSTm,\Gpost}  &  \Argdef{daswp}\\
\t1 \Gpost\rN{a}{0} &\\
 \end{argue}
 So:
 \begin{argue}
 \Infection & \Argcalc{\uparrow} \\
\t1  \Gpost\rN{a}{0-a}  \neq    \Gpost\rN{a}{0} &\\
 \end{argue}
 
 If $0-a \neq 0$  which means $ a \neq 0 $ then $\AMDm$ will be killed weakly.

  \begin{argue}
   \underline{{\bf Propagation}}: & \Argdef{ppgdefnif}\\
\Propagationgif(\STju,\STjm)=  \\
 \Propagationnif= & \Argdef{isevendef} \\
 \wpre{\STju \compose \isEvenPRja \compose \PRa,\Gpost}  \neq \wpre{\STjm \compose \isEvenPRja \compose \PRa,\Gpost} & \Arglaw{\lNcP}\\
 \wpre{\STju \compose \PRa,\Gpost}  \neq \wpre{\STjm \compose \PRa,\Gpost} &\Argdef{dscwp}\\
  \wpre{\STju, \wpre{\PRa,\Gpost}}  \neq \wpre{\STjm ,\wpre{ \PRa,\Gpost}} & \\
  \end{argue}
  \small
   \begin{argue}
  \wpre{\PRa,\Gpost} = & \Argdef{isevendef}\\ 
\t1    \wpre{\isEvenPRa,\Gpost} = \\
& \Argdef{difwp}\\
\t1  (\floor{\frac{a}{2}} = \frac{a}{2} \lor \floor{\frac{a}{2}} \neq \frac{a}{2}) \land \\
\t1 (\floor{\frac{a}{2}} = \frac{a}{2} \implies \wpre{\returnst(\truest),\Gpost}) \land\\
\t1  ( \floor{\frac{a}{2}} \neq \frac{a}{2} \implies \wpre{\returnst(\falsest),\Gpost}) = & \Arglaw{\lAonA},\Argdef{drtwp}\\
\t1  \truest \land (\floor{\frac{a}{2}} = \frac{a}{2} \implies \Gpost\rn{\AMD}{\truest}) \land  (\floor{\frac{a}{2}} \neq \frac{a}{2} \implies \Gpost\rn{\AMD}{\falsest}) = \\
\t1  (\floor{\frac{a}{2}} = \frac{a}{2} \implies \Gpost\rn{\AMD}{\truest}) \land  (\floor{\frac{a}{2}} \neq \frac{a}{2} \implies \Gpost\rn{\AMD}{\falsest})  \\
\end{argue}
  \begin{argue}
   \wpre{\STju, \wpre{\PRa,\Gpost}} = &\Argcalc{\uparrow}, \Argdef{isevendef}\\ 
\t1   \wpre{\isEvenPRjSTu, (\floor{\frac{a}{2}} = \frac{a}{2} \implies \Gpost\rn{\AMDu}{\truest}) \land  (\floor{\frac{a}{2}} \neq \frac{a}{2} \implies \Gpost\rn{\AMDu}{\falsest}) } = &\\
&  \Argdef{daswp}\\
\t1 ((\floor{\frac{a}{2}} = \frac{a}{2} \implies \Gpost\rn{\AMDu}{\truest}) \land  (\floor{\frac{a}{2}} \neq \frac{a}{2} \implies \Gpost\rn{\AMDu}{\falsest}))\rN{a}{0-a}=\\
\t1 ((\floor{\frac{0-a}{2}} = \frac{0-a}{2} \implies \Gpost\rn{\AMDu}{\truest}) \land  (\floor{\frac{0-a}{2}} \neq \frac{0-a}{2} \implies \Gpost\rn{\AMDu}{\falsest}))\\
\end{argue}

  \begin{argue}
   \wpre{\STjm, \wpre{\PRa,\Gpost}} = & \Argcalc{\uparrow}, \Argdef{isevendef}\\ 
\t1   \wpre{\isEvenPRjSTm, (\floor{\frac{a}{2}} = \frac{a}{2} \implies \Gpost\rn{\AMDm}{\truest}) \land  (\floor{\frac{a}{2}} \neq \frac{a}{2} \implies \Gpost\rn{\AMDm}{\falsest}) } = &\\
&  \Argdef{daswp}\\
\t1 ((\floor{\frac{a}{2}} = \frac{a}{2} \implies \Gpost\rn{\AMDm}{\truest}) \land  (\floor{\frac{a}{2}} \neq \frac{a}{2} \implies \Gpost\rn{\AMDm}{\falsest}))\rN{a}{0}=\\
\t1 ((\floor{\frac{0}{2}} = \frac{0}{2} \implies \Gpost\rn{\AMDm}{\truest}) \land  (\floor{\frac{0}{2}} \neq \frac{0}{2} \implies \Gpost\rn{\AMDm}{\falsest}))\\
\end{argue}

  \begin{argue}  
 \Propagationgif \defs  \Propagationnif &\Argcalc{\uparrow}\\
\t1 ((\floor{\frac{0-a}{2}} = \frac{0-a}{2} \implies \Gpost\rn{\AMDu}{\truest}) \land  (\floor{\frac{0-a}{2}} \neq \frac{0-a}{2} \implies \Gpost\rn{\AMDu}{\falsest})) \neq  \\
\t1 ((\floor{\frac{0}{2}} = \frac{0}{2} \implies \Gpost\rn{\AMDm}{\truest}) \land  (\floor{\frac{0}{2}} \neq \frac{0}{2} \implies \Gpost\rn{\AMDm}{\falsest}))\\
  \end{argue}
  \normalsize
  This shows that if $a$ is even  then the result of $\AMDm$ is the same as the result of $\AMDu$, but when $a$ is not even  then these two results are different. As a result, the propagation condition is $\floor{\frac{0-a}{2}} \neq \frac{0-a}{2}$(``$a$ is not even").

 \begin{argue}
  \underline{{\bf Full~ Test~ Specification}}:&  \Argdef{ftsdefripnif}\\
\FTSripgif(\STju,\STjm) = \\
 \FTSripnif = & \Argcalc{\uparrow}\\
 a < 0 \land a \neq 0 \land   \floor{\frac{0-a}{2}} \neq \frac{0-a}{2}
 \end{argue}
 
  This shows that for $a < 0$ if $\floor{\frac{0-a}{2}} = \frac{(0-a)}{2}$($a$ is even)  then the result of $\AMDm$ is the same as the result of $\AMDu$, but when $a$ is not even  then these two results are different. As a result, the full test specification  is $a <0 \land \floor{\frac{0-a}{2}} \neq \frac{0-a}{2}$(``$a$ is negative and not even") and $\AMDm$ is killed strongly.
 

%% file: zp0404.tex
\subsection{Chkdig Program: Guarded Command Modification}
\subsubsection{Definitions}
The original code in which the mutated statements are marked by prime, discussed in Example \ref{exampchkdig}, is as follow:

\pchkdigvb

The translation of this program and its mutated version for Mutant 1 to Dijkestra Guarded Commands are as follow (Example \ref{exampchkdiggc}):
 \renewcommand{\AMD}{Chkdig}
\renewcommand{\AMDu}{\undertest{\AMD}}
\renewcommand{\AMDma}{\mutateda{\AMD}}
\[\AMDu\defs \begin{array}{l} \pchkdig \end{array} \]
\[\AMDma \defs\begin{array}{l} \pchkdigma \end{array} \]

\newcommand{\ChkdigPRb}{\left\{\begin{array}{l} \pchkdigba \end{array}\right\}}

\renewcommand{\ChkdigSTu}{\left\{\begin{array}{l} \pchkdigstua \end{array}\right\}}
\renewcommand{\ChkdigSTuif}{\left\{\begin{array}{l} \pchkdigstuifa \end{array}\right\}}
\renewcommand{\ChkdigPRjb}{\pchkdigprjba}
\renewcommand{\ChkdigPRjSTu}{\pchkdigprjstua}
\renewcommand{\ChkdigPRja}{\pchkdigprjaa}
\renewcommand{\ChkdigPRj}{\left\{\begin{array}{l} \ChkdigPRjb\\ \ChkdigPRjSTu\\ \ChkdigPRja \end{array}\right\}}
\renewcommand{\ChkdigGj}{\pchkdiggja }
\newcommand{\ChkdigPRa}{\left\{\begin{array}{l} \pchkdigaa \end{array}\right\}}
\newcommand{\ChkdigSTm}{\left\{\begin{array}{l} \pchkdigstma \end{array}\right\} }
\newcommand{\ChkdigSTmif}{\left\{\begin{array}{l} \pchkdigstmifa \end{array}\right\} }
\newcommand{\ChkdigPRjSTm}{\pchkdigprjstma}
\renewcommand{\ChkdigGdo}{\pchkdiggja}
\renewcommand{\ChkdigGdoj}{\pchkdiggja}
\renewcommand{\ChkdigIdo}{0 \leq r \leq a \land (d=0 \lor d=\lastd{r}) }
\renewcommand{\ChkdigVdo}{r}

  The underlined statement is the mutated statement. This is an example  of a guarded command modification. To calculate the RIP conditions,  the above programs and what are given in Definitions \ref{putdefn} and  \ref{mutdefn} must be compared. So:\\
\begin{Definition} 
\label{chkdigdefa}
\begin{syntax}
\PRb & : &  \ChkdigPRb  \\
\STu & : & \ChkdigSTu   \\
\PRjb &:& \ChkdigPRjb\\
\STju & : & \ChkdigPRjSTu   \\
\PRja &:& \ChkdigPRja\\
\PRa & : & \ChkdigPRa \\
\STm & : &  \ChkdigSTm\\
\STjm & : &  \ChkdigPRjSTm\\
\Gj  &:&   \ChkdigGj\\
\Gdo   &:& \ChkdigGdo\\
\Gdoj   &:& \ChkdigGdoj\\
\Ido   &:& \ChkdigIdo\\
\Vdo   &:& \ChkdigVdo\\
\end{syntax}
\end{Definition}

\subsubsection{RIP Calculations}

\begin{argue}
\underline{{\bf Reachability}}: &\Argdef{rchdefn}\\
\Reachabilityg(\STju)=  \Reachabilityn   & \Argdef{chkdigdefa}\\
\Reachabilityg(\ChkdigPRjSTu)=     & \\
\t1 \reachc{\ChkdigPRb}  \land  \ChkdigGj \land  \reachc{\ChkdigPRjb} = \\
& \Argdef{dscrc}\\
\t2 \reachc{s := 1}  \land & \Argdef{dasrc}\\
\t2 \reachc{\ifst (b\leq 1 \lor b > 10 \lor a < 0)  \rightarrow \returnst(s) \\
\t3 \elsest (b > 1 \land b \leq 10 \land a \geq 0) \rightarrow \skipst}  \land & \Argdef{difrc}\\   
\t2 \reachc{s:= 2} \land
 \reachc{ r :=a} \land
\reachc{ d:=0} \land & \Argdef{dasrc}\\ 
\t2 \ChkdigGj \land \\
\t2 \reachc{t := r} \land 
 \reachc{ r := \floor{\frac{ t}{10}}}= & \Argdef{dasrc}\\
\t2 \truest  \land & \\
\t2 (((b\leq 1 \lor b > 10 \lor a < 0)\land \reachc{\returnst(s)}) \lor \\
\t3 ((b > 1 \land b \leq 10 \land a \geq 0) \land \reachc{\skipst}))  \land & \Argdefs{drtrc}{dskrc}\\   
\t2 \truest \land
 \truest \land
\truest \land & \Argdef{dasrc}\\ 
\t2 \ChkdigGj \land \\
\t2 \truest \land 
 \truest= & \Argdef{dasrc}\\
\t2 (((b\leq 1 \lor b > 10 \lor a < 0)\land \falsest ) \lor 
 ((b > 1 \land b \leq 10 \land a \geq 0) \land \truest))  \land  
 \ChkdigGj = \\
\t2 (\falsest \lor 
 ((b > 1 \land b \leq 10 \land a \geq 0) \land \truest))  \land  
 \ChkdigGj = \\
\t2 (b > 1 \land b \leq 10 \land a \geq 0)  \land  \ChkdigGj  \\
\end{argue}
The reachability condition shows that the base($b$) must be between $1$ and $10$ and the input($a$) must be greater than or equal to zero. The last conjunct means that the loop condition must be true at  the begining of iterations of loop. It is equivalent to $a>0 \land 0 < b$.

\begin{argue}
\underline{{\bf Infection}}: &  \Argdef{infdefn}\\
\Infectiong(\STju,\STjm)= \Infectionn & \\
\wpre{\STju,\Gpost}=  & \Argdef{chkdigdefa}\\
\t1  \wpre{\ChkdigPRjSTu,\Gpost}= &  \Argdef{daswp} \\
\t1  \Gpost\rN{d}{t-r*10} \\
 \end{argue}
 
 \begin{argue}
 \wpre{\STjm,\Gpost}=  & \Argdef{chkdigdefa}\\
\t1  \wpre{\ChkdigPRjSTm,\Gpost}= &  \Argdef{daswp} \\
\t1  \Gpost\rN{d}{t+r*10}= &  \Argdef{daswp} \\
\end{argue}
\begin{argue}
\underline{{\bf Infection}}: &  \Argdef{infdefn}\\
\Infectiong(\STju,\STjm)= \Infectionn & \\
\t1  \Gpost\rN{d}{t-r*10} \neq   \Gpost\rN{d}{t+r*10}
 \end{argue}

The infection condition is $\truest$ if $r>0$ which means $a \geq 10$.

    \begin{argue}
    \underline{{\bf Propagation}}: & \Argdef{ppgdefndo}\\   
 \Propagationgdo(\STju,\STjm)=  \Propagationndo &\\
 &   \Argdef{dscwp}\\
\wpre{\STu, \wpre{\PRa,\Gpost}}  \neq \wpre{ \STm ,\wpre{ \PRa,\Gpost}}=\\
\end{argue}
\begin{argue}
\wpre{\PRa,\Gpost}=  & \Argdef{chkdigdefa}\\
 \t1 \wpre{\ChkdigPRa,\Gpost}= & \Argdef{dscwp}\\
\t1  \wpre{\ifst ((d  < b)  \land  (r = 0)) \rightarrow s :=3 \elsest ((d \geq b) \lor (r \neq 0)) \rightarrow \skipst, \wpre{\returnst(s),\Gpost}} = & \\
& \Argdef{drtwp}\\
\t1  \wpre{\ifst ((d  < b)  \land  (r = 0)) \rightarrow s :=3 \elsest ((d \geq b) \lor (r \neq 0)) \rightarrow \skipst, \Gpost\rN{\AMDu}{s}} = &\\
&  \Argdef{difwp}\\
\t1  (((d  < b)  \land  (r = 0)) \lor ((d \geq b) \lor (r \neq 0))) \land & \Arglaw{\lAonA}\\
\t1 (((d  < b)  \land  (r = 0)) \implies \wpre{ s :=3,\Gpost\rN{\AMDu}{s}}) \land  & \Argdef{daswp}\\
\t1 (((d \geq b) \lor (r \neq 0)) \implies \wpre{\skipst, \Gpost\rN{\AMDu}{s}}) = & \Argdef{dskwp}\\
\t1\truest  \land & \\
\t1 (((d  < b)  \land  (r = 0)) \implies \Gpost\rN{\AMDu}{s}\rN{s}{3}) \land  \\
\t1 (((d \geq b) \lor (r \neq 0)) \implies \Gpost\rN{\AMDu}{s}) = & \\
\t1 ((d  < b  \land  r = 0) \implies \Gpost\rN{\AMDu}{s}\rN{s}{3}) \land\\
\t1   ((d \geq b \lor r \neq 0) \implies \Gpost\rN{\AMDu}{s})= & \\
\t1 \underbrace{((d  < b  \land  r = 0) \implies \Gpost\rN{\AMDu}{3}) \land  ((d \geq b \lor r \neq 0) \implies \Gpost\rN{\AMDu}{s})}_{\Gpostb} & \\
\end{argue}
\newcommand{\AGpostb}{((d  < b  \land  r = 0) \implies \Gpost\rN{\AMDu}{3}) \land  ((d \geq b \lor r \neq 0) \implies \Gpost\rN{\AMDu}{s})}
\begin{argue}
\wpre{\STu, \wpre{\PRa,\Gpost}}= & \Argdef{chkdigdefa},\Argcalc{\uparrow}\\
\t1 \wpre{\ChkdigSTu,\Gpostb } =& \Argdef{ddowp},\Argcalc{k=1}\\
\t1 \wpre{\ChkdigSTuif, \DHp{0}{\Gpostb} }  & \\
\t1 \wpre{\ChkdigSTuif, \DHp{0}{\Gpostb} }  =& \Argdef{difwp}\\
\t1 (\ChkdigGdo \lor \lnot \ChkdigGdo) \land  & \Arglaw{\lAonA}\\
\t1 \ChkdigGdo \implies  \wpre{\ChkdigPRjb\compose \ChkdigPRjSTu\compose\ChkdigPRja,\DHp{0}{\Gpostb} } \land & \Argdef{dscwp}\\
\t1 \lnot \ChkdigGdo \implies  \wpre{\skipst,\DHp{0}{\Gpostb} } =& \Argdef{dskwp}\\
\t1 \truest \land  &\\
\t1 \ChkdigGdo \implies  \wpre{ t := r, \wpre{ r := \floor{\frac{ t}{10}}, \wpre{\ChkdigPRjSTu, \wpre{\ChkdigPRja,\DHp{0}{\Gpostb} }}}} =\\
\t1 \lnot \ChkdigGdo \implies  \DHp{0}{\Gpostb}  =& \Argdef{dnlwp} \\
\t1 \ChkdigGdo \implies  \wpre{ t := r, \wpre{ r := \floor{\frac{ t}{10}}, \wpre{\ChkdigPRjSTu,\DHp{0}{\Gpostb}  }}} \land &\\
\t1 \lnot \ChkdigGdo \implies  \DHp{0}{\Gpostb}  =&  \Argdef{daswp} \\
\t1 \ChkdigGdo \implies  \wpre{ t := r, \wpre{ r := \floor{\frac{ t}{10}},   \lba\DHp{0}{\Gpostb}  \rba\rN{d}{t-r*10} }} \land & \\
\t1 \lnot \ChkdigGdo \implies  \DHp{0}{\Gpostb}  =&  \Argdef{daswp} \\
\t1 \ChkdigGdo \implies  \wpre{ t := r,   \lbb \lba\DHp{0}{\Gpostb} \rba\rN{d}{t-r*10}\rbb\rN{ r}{\floor{\frac{ t}{10}}}} \land & \\
\t1 \lnot \ChkdigGdo \implies  \DHp{0}{\Gpostb}  = & \Argdef{daswp} \\
\t1 \ChkdigGdo \implies   \lbc \lbb \lba\DHp{0}{\Gpostb} \rba\rN{d}{t-r*10}\rbb\rN{ r}{\floor{\frac{ t}{10}}}\rbc \rN{ t }{r}  \land & \\
\t1 \lnot \ChkdigGdo \implies  \DHp{0}{\Gpostb}  =& \\
\end{argue}
We know that:
\begin{argue}
\DHp{0}{\Gpostb}=\Gpostb \land \lnot \ChkdigGdo & \Argdef{ddowp}\\
\Gpostb=\AGpostb & \Argcalc{\uparrow}\\
\DHp{0}{\Gpostb}=\AGpostb    \land \lnot \ChkdigGdo
\end{argue}
So we have:
\begin{argue}
\t1 \ChkdigGdo \implies\\
\t2    \lbc \lbb \lba\AGpostb    \land \lnot \ChkdigGdo \rba\\
\t3 \rN{d}{t-r*10}\rbb\rN{ r}{\floor{\frac{ t}{10}}}\rbc \rN{ t }{r}  \land & \\
\t1 \lnot \ChkdigGdo \implies  \DHp{0}{\Gpostb}  =& \\
\t1 \ChkdigGdo \implies\\
\t2  (((r-\floor{\frac{ r}{10}}*10) < b  \land  \floor{\frac{ r}{10}} = 0) \implies \Gpost\rN{\AMDu}{3}) \land \\
\t2  (((r-\floor{\frac{ r}{10}}*10) \geq b \lor \floor{\frac{ r}{10}} \neq 0) \implies \Gpost\rN{\AMDu}{s})   \land \lnot ( \floor{\frac{ r}{10}} > 0 \land (r-\floor{\frac{ r}{10}}*10) < b)   \land & \\
\t1 \lnot \ChkdigGdo \implies \\
\t2  ((d  < b  \land  r = 0) \implies \Gpost\rN{\AMDu}{3}) \land  ((d \geq b \lor r \neq 0) \implies \Gpost\rN{\AMDu}{s})   \land \lnot \ChkdigGdo  =& \\
\end{argue}
\begin{argue}
\wpre{\STm, \wpre{\PRa,\Gpost}}=\\
\t1 \wpre{\ChkdigSTm,\Gpostb } =\\
& \Argcalc{\uparrow}\\
\t1 \ChkdigGdo \implies\\
\t2  (((r+\floor{\frac{ r}{10}}*10) < b  \land  \floor{\frac{ r}{10}} = 0) \implies \Gpost\rN{\AMDma}{3}) \land \\
\t2  (((r+\floor{\frac{ r}{10}}*10) \geq b \lor \floor{\frac{ r}{10}} \neq 0) \implies \Gpost\rN{\AMDma}{s})   \land \lnot ( \floor{\frac{ r}{10}} > 0 \land (r-\floor{\frac{ r}{10}}*10) < b)   \land & \\
\t1 \lnot \ChkdigGdo \implies \\
\t2 ((d  < b  \land  r = 0) \implies \Gpost\rN{\AMDma}{3}) \land  ((d \geq b \lor r \neq 0) \implies \Gpost\rN{\AMDma}{s})    \land \lnot \ChkdigGdo  =& \\
\end{argue}
\newcommand{\AGpostc}{\lnot\ChkdigGdo \land \AGpostb}
   \begin{argue}
    \underline{{\bf Propagation}}: & \Argdef{ppgdefndo}\\   
 \Propagationgdo(\STju,\STjm)=  \Propagationndo &\\
\t1 \ChkdigGdo \implies\\
\t2  (((r-\floor{\frac{ r}{10}}*10) < b  \land  \floor{\frac{ r}{10}} = 0) \implies \Gpost\rN{\AMDu}{3}) \land \\
\t2  (((r-\floor{\frac{ r}{10}}*10) \geq b \lor \floor{\frac{ r}{10}} \neq 0) \implies \Gpost\rN{\AMDu}{s})   \land \lnot ( \floor{\frac{ r}{10}} > 0 \land (r-\floor{\frac{ r}{10}}*10) < b)   \land & \\
\t1 \lnot \ChkdigGdo \implies \\
\t2  ((d  < b  \land  r = 0) \implies \Gpost\rN{\AMDu}{3}) \land  ((d \geq b \lor r \neq 0) \implies \Gpost\rN{\AMDu}{s})   \land \lnot \ChkdigGdo  \neq \\
\t1 \ChkdigGdo \implies\\
\t2  (((r+\floor{\frac{ r}{10}}*10) < b  \land  \floor{\frac{ r}{10}} = 0) \implies \Gpost\rN{\AMDma}{3}) \land \\
\t2  (((r+\floor{\frac{ r}{10}}*10) \geq b \lor \floor{\frac{ r}{10}} \neq 0) \implies \Gpost\rN{\AMDma}{s})   \land \lnot ( \floor{\frac{ r}{10}} > 0 \land (r+\floor{\frac{ r}{10}}*10) < b)   \land & \\
\t1 \lnot \ChkdigGdo \implies \\
\t2 ((d  < b  \land  r = 0) \implies \Gpost\rN{\AMDma}{3}) \land  ((d \geq b \lor r \neq 0) \implies \Gpost\rN{\AMDma}{s})    \land \lnot \ChkdigGdo
\end{argue}

To analyze the above propagation condition, some cases are distinguished as follow:
\begin{itemize}
\item   $\AMDu$ and $\AMDma$ do not enter the loop: Since $\ChkdigGdo=\falsest$ then both sides of propagation condition are $\truest$, as a result propagation condition is $\falsest$.
\item  $\AMDu$ and $\AMDma$ enter the loop: $\lnot \ChkdigGdo$ means after one iteration($k=1$) the guard must become $\falsest$ and  it means $d \geq b$ or $r =0$ must be $\truest$. Three cases must be considered:
\begin{itemize}
\item $d \geq b\equiv\truest$ and $r =0\equiv \falsest$: This means $a \geq 10$ and the results of both programs are identical. So, propagation condition is $\falsest$.
\item $d \geq b\equiv\falsest$ and  $r =0\equiv \truest$: This means $0<a <10$ and $\alld{a}{b}$ and the results of both programs are identical. So, propagation condition is $\falsest$.
\item $d \geq b\equiv\falsest$ and  $r =0\equiv \falsest$: This means $a \geq10$ and $\alld{a}{b}$.  Only in this case  the result of $\AMDu$ and the result of $\AMDma$ are different. So, propagation condition is $\truest$.
\end{itemize} 
\end{itemize}
The conclusion is the propagation condition is $a \geq 10 \land \alld{a}{b}$.
  \begin{argue}
   \underline{{\bf Full~ Test~ Specification}}: &  \Argdef{ftsdefripndo}\\
 \FTSripgdo(\STju,\STjm) = \\
 \FTSripndo = & \Argcalc{\uparrow}\\
\t2  (b > 1 \land b \leq 10 \land a \geq 0)  \land \\
\t2 a \geq 10    \land \\
\t2 a \geq 10 \land \alld{a}{b} =\\
\t2  b > 1 \land b \leq 10 \land a \geq 10  \land \alld{a}{b}
 \end{argue}
 The full test specification shows that the test case which can kill the mutant($\AMDma$) strongly is the one with base greater than 1 and less than or equal to 10($b > 1 \land b \leq 10$) and  input($a$) must be greater than $10$ and all the digits of $a$ must be less than the base($b$). In this case the result of $\AMDu$  is 3 but the result of $\AMDma$ is 2. For instance, assume, $a$ is set to $15$ and the base is set to $6$ then the result of $\AMDu$ is $3$ but the result of $\AMDma$ is $2$. But, if $a$ is set to $15$ and the base is set to $4$ then the result of $\AMDu$ is $2$ and the result of $\AMDma$ is $2$ also, despite that infection occurs but not propagated.

%% file: zp0405.tex
\subsection{Chkdig Program: Statement Modification}
\subsubsection{Definitions}
The original code in which the mutated statements are marked by prime, discussed in Example \ref{exampchkdig}, is as follow:

\pchkdigvb

The translation of this program and its mutated version for Mutant 2 to Dijkestra Guarded Commands are (Example \ref{exampchkdiggc}):
 \renewcommand{\AMD}{Chkdig}
\renewcommand{\AMDu}{\undertest{\AMD}}
\renewcommand{\AMDmb}{\mutatedb{\AMD}}
\[\AMDu\defs \begin{array}{l} \pchkdig \end{array} \]
\[\AMDmb \defs\begin{array}{l} \pchkdigmb \end{array} \]

\renewcommand{\ChkdigPRb}{\left\{\begin{array}{l} \pchkdigbb \end{array}\right\}}

\renewcommand{\ChkdigSTu}{\left\{\begin{array}{l} \pchkdigstub \end{array}\right\}}
\renewcommand{\ChkdigPRa}{\pchkdigab}
\renewcommand{\ChkdigSTm}{\left\{\begin{array}{l} \pchkdigstmb \end{array}\right\} }

  The underlined statement is the mutated statement. This is again a statement modification. To calculate the RIP conditions, the above programs and what are given in Definitions \ref{putdef} and \ref{mutdef} must be compared. So:\\
\begin{Definition} 
\label{chkdigdefb}
\begin{syntax}
\PRb & : &  \ChkdigPRb  \\
\STu & : & \ChkdigSTu   \\
\PRa & : & \ChkdigPRa \\
\STm & : &  \ChkdigSTm\\
\Gdo   &:& \ChkdigGdo\\
\end{syntax}
\end{Definition}

\subsubsection{RIP Calculations}

\begin{argue}
\underline{{\bf Reachability}}: &\Argdef{rchdef}\\
\Reachabilitys(\STu)=  \Reachability   & \Argdef{chkdigdefb}\\
\Reachabilitys(\ChkdigSTu)=    & \\
\t1 \reachc{\ChkdigPRb} = & \Argdef{dnlrc}\\
\t2  \reachc{s := 1} \land  & \Argdef{dasrc}\\
\t2  \reachc{ \ifst (b\leq 1 \lor b > 10 \lor a < 0)  \rightarrow \returnst(s)   \elsest (b > 1 \land b \leq 10 \land a \geq 0) \rightarrow \skipst} \land  & \Argdef{difrc}\\
 \t2  \reachc{s:= 2}  \land   \reachc{r :=a} \land    \reachc{ d:=0} \land & \Argdef{dasrc}\\
 \t2 \reachc{\dost  ( r > 0 \land d < b) \rightarrow  t := r\compose r := \floor{\frac{ t}{10}} \compose d := t-r*10 \odst}= & \Argdef{ddorc}\\ 
\t2  \truest \land  \\
\t2  (((b\leq 1 \lor b > 10 \lor a < 0)  \land \reachc{\returnst(s)})  \lor\\
\t2  ((b > 1 \land b \leq 10 \land a \geq 0) \land \reachc{ \skipst})) \land & \Argdefs{drtrc}{dskrc}\\
 \t2 \truest  \land \truest \land  \truest \land &\\
 \t2 \lnot( r > 0 \land d < b)= &\\
\t2  (((b\leq 1 \lor b > 10 \lor a < 0)  \land \falsest)  \lor  ((b > 1 \land b \leq 10 \land a \geq 0) \land \truest)) \land  \lnot( r > 0 \land d < b) = &\\
\t2  (\falsest  \lor (b > 1 \land b \leq 10 \land a \geq 0)) \land  \lnot( r > 0 \land d < b) = &\\
\t2  (b > 1 \land b \leq 10 \land a \geq 0) \land ( r \leq  0 \lor d \geq b)  &\\
\end{argue}
The reachability condition shows that the base($b$) must be between $1$ and $10$ and the input($a$) must be greater than or equal to zero. The last conjunct means that the loop condition must be false, from the begining or after some iterations. 

\begin{argue}
\underline{{\bf Infection}}: &  \Argdef{infdef}\\
\Infections(\STu,\STm)= \Infection & \\
\wpre{\STu,\Gpost}=  & \Argdef{chkdigdefb}\\
\t1  \wpre{\ChkdigSTu,\Gpost}= &  \Argdef{difwp} \\
\t1  (((d  < b)  \land  (r = 0)) \lor ((d \geq b) \lor (r \neq 0))) \land \\
\t1  ((d  < b)  \land  (r = 0)) \implies \wpre{s :=3, \Gpost} \land \\
\t1  ((d \geq b) \lor (r \neq 0)) \implies \wpre{\skipst,\Gpost} =& \Arglaw{\lAonA}, \Argdefs{daswp}{dskwp} \\
\t1  ((d  < b)  \land  (r = 0)) \implies \Gpost\rN{s}{3} \land & \Arglaw{\lAaBiC}\\
\t1 ((d \geq b) \lor (r \neq 0)) \implies \Gpost =& \Arglaw{\lAoBiC}\\
\t1  ((d  < b \implies \Gpost\rN{s}{3})  \lor   (r = 0 \implies \Gpost\rN{s}{3})) \land  ((d \geq b \implies \Gpost)  \land (r \neq 0 \implies \Gpost)) =& \\
\t1  ((d  < b \implies \Gpost\rN{s}{3}) \land  (d \geq b \implies \Gpost)  \land (r \neq 0 \implies \Gpost)) \lor  & \\
\t1 ((r = 0 \implies \Gpost\rN{s}{3}) \land  (d \geq b \implies \Gpost)  \land (r \neq 0 \implies \Gpost)) =\\
& \Arglaw{\lAiBanAiC}\\
\t1  (((d  < b \land \Gpost\rN{s}{3}) \lor  (d \geq b \land \Gpost))  \land (r \neq 0 \implies \Gpost)) \lor  & \\
\t1 (((r = 0 \land \Gpost\rN{s}{3}) \lor (r \neq 0 \land \Gpost))\land  (d \geq b \implies \Gpost)  )= &\\
\t1 (d  < b \land \Gpost\rN{s}{3} \land (r \neq 0 \implies \Gpost))\lor  (d \geq b \land \Gpost  \land (r \neq 0 \implies \Gpost)) \lor  & \\
\t1 (r = 0 \land \Gpost\rN{s}{3} \land  (d \geq b \implies \Gpost))  \lor (r \neq 0 \land \Gpost \land  (d \geq b \implies \Gpost))= & \\
& \Arglaw{\lAiB}\\
\t1 ((d  < b \land \Gpost\rN{s}{3}) \land (r = 0 \lor \Gpost))\lor  ((d \geq b \land \Gpost)  \land (r = 0 \lor \Gpost)) \lor  & \\
\t1 ((r = 0 \land \Gpost\rN{s}{3}) \land  (d < b \lor \Gpost))  \lor ((r \neq 0 \land \Gpost) \land  (d < b \lor \Gpost))= &\\
 \end{argue}
 \begin{argue}
\t1 (d  < b \land r = 0 \land \Gpost\rN{s}{3} ) \lor\\
\t1 (d  < b \land \Gpost\rN{s}{3} \land \Gpost)\lor & \Arglaw{\Gpost\rN{s}{3} \land \Gpost= \Gpost\rN{s}{3}}\\
\t1 (d \geq b \land r = 0 \land \Gpost  ) \lor\\
 \t1 (d \geq b \land \Gpost  \land\Gpost) \lor  & \\
\t1 (d < b \land r = 0 \land \Gpost\rN{s}{3} ) \lor\\
\t1 (r = 0 \land \Gpost\rN{s}{3}\land  \Gpost)  \lor & \Arglaw{\Gpost\rN{s}{3} \land \Gpost= \Gpost\rN{s}{3}}\\
\t1 (d < b \land r \neq 0 \land \Gpost ) \lor\\
\t1 (r \neq 0 \land \Gpost \land  \Gpost)= &\\
\t1 (d  < b  \land r = 0 \land \Gpost\rN{s}{3}) \lor\\
\t1 (d < b \land r \neq 0 \land \Gpost  ) \lor\\
\t1 (d  < b \land \Gpost\rN{s}{3})\lor\\
\t1 (d \geq b \land r = 0 \land \Gpost  ) \lor\\
 \t1 (d \geq b \land \Gpost ) \lor  & \\
\t1 (r = 0 \land \Gpost\rN{s}{3})  \lor\\
\t1 (r \neq 0 \land \Gpost) &\\
 \end{argue}
 \newcommand{\pra}{a}
\newcommand{\pran}{\bar{a}}
 \newcommand{\prb}{b}
 \newcommand{\prbn}{\bar{b}}
 \newcommand{\prc}{c}
  \newcommand{\prcn}{\bar{c}}
 \newcommand{\prd}{d}
 \newcommand{\prdn}{\bar{d}}
 
  \newcommand{\ppra}{d<b}
\newcommand{\ppran}{d \geq b}
 \newcommand{\pprb}{r=0}
 \newcommand{\pprbn}{r\neq 0}
 \newcommand{\pprc}{\Gpost}
  \newcommand{\pprcn}{\lnot\Gpost}
 \newcommand{\pprd}{\Gpost\rN{s}{3}}
 \newcommand{\pprdn}{\lnot\Gpost\rN{s}{3}}


Assuming $\pra:\ppra$, $\prb:\pprb$, $\prc:\pprc$, and $\prd:\pprd$, and using multiplication instead of conjunction and addition instead of disjunction,  the above predicate can be rewritten as:

\begin{argue}
\t1 \pra\prb\prd+  
\pra\prbn\prc+ 
\pra\prd+ 
\pran\prb\prc+
\pran\prc+
\prb\prd+ 
\prbn\prc=
\end{argue}

\begin{figure}[h]
\centering
\scriptsize
\roundbox{
\begin{minipage}{7cm}

\karnaughmap{4}{$f$}%
{{$d$}{$b$}{$c$}{$a$}}%
{%
00110010%
01111111
}%
{%
\textcolor{Red}{
\put(2,1){\oval(1.9,1.9)[]}}
\textcolor{Blue}{
\put(.85,2){\oval(1.9,1.9)[]}}
\textcolor{Green}{
\put(2.70,1){\oval(1.9,1.9)[]}}
\textcolor{Black}{
\put(-.5,2){\oval(1.9,1.9)[r]}
\put(3.5,2){\oval(1.9,1.9)[l]}}
}
\end{minipage}
}
\normalsize
\caption{\label{karnoafig}
Karno Map for $ \pra\prb\prd+  \pra\prbn\prc+ \pra\prd+ \pran\prb\prc+\pran\prc+\prb\prd+ \prbn\prc$ }
\end{figure}
Using Karno Map (Figure \ref{karnoafig}),  this predicate can be simplified as:
\begin{argue}
\t1 \pra\prd+\pran\prc +\prb\prd +\prbn\prc
\end{argue}

 Which is:
\begin{argue}
\t1 (\ppra \land \pprd) \lor (\ppran \land \pprc)  \lor (\pprb \land \pprd)  \lor (\pprbn \land \pprc)
\end{argue}

 \begin{argue}
 \wpre{\STm,\Gpost}=  & \Argdef{chkdigdefb}\\
\t1  \wpre{\ChkdigSTm,\Gpost}= &  \Argdef{difwp} \\
\t1  (((d  < b)  \lor  (r = 0)) \lor ((d \geq b) \land (r \neq 0))) \land \\
\t1  ((d  < b)  \lor  (r = 0)) \implies \wpre{s :=3, \Gpost} \land \\
\t1  ((d \geq b) \land (r \neq 0)) \implies \wpre{\skipst,\Gpost} =& \Arglaw{\lAonA}, \Argdefs{daswp}{dskwp} \\
\t1  ((d  < b)  \lor  (r = 0)) \implies \Gpost\rN{s}{3} \land & \Arglaw{\lAoBiC}\\
\t1   ((d \geq b) \land (r \neq 0)) \implies \Gpost= & \Arglaw{\lAaBiC} \\
\t1  ((d  < b  \implies \Gpost\rN{s}{3})  \land  (r = 0 \implies \Gpost\rN{s}{3})) \land & \\
\t1   ((d \geq b  \implies \Gpost) \lor (r \neq 0 \implies \Gpost))= &  \\
\t1  ((d  < b  \implies \Gpost\rN{s}{3})  \land  (r = 0 \implies \Gpost\rN{s}{3}) \land   (d \geq b  \implies \Gpost)) \lor\\
\t1   ((d  < b  \implies \Gpost\rN{s}{3})  \land  (r = 0 \implies \Gpost\rN{s}{3}) \land (r \neq 0 \implies \Gpost)) &  \\
& \Arglaw{\lAiBanAiC}\\
\t1  (((d  < b  \land\Gpost\rN{s}{3})  \lor   (d \geq b  \land \Gpost))  \land  (r = 0 \implies \Gpost\rN{s}{3})) \lor\\
\t1   (((r = 0 \land \Gpost\rN{s}{3}) \lor (r \neq 0 \land \Gpost)) \land (d  < b  \implies \Gpost\rN{s}{3}) )=   &  \\
\t1  (d  < b  \land\Gpost\rN{s}{3}  \land  (r = 0 \implies \Gpost\rN{s}{3}))  \lor   (d \geq b  \land \Gpost  \land  (r = 0 \implies \Gpost\rN{s}{3})) \lor\\
\t1   (r = 0 \land \Gpost\rN{s}{3} \land (d  < b  \implies \Gpost\rN{s}{3})) \lor (r \neq 0 \land \Gpost \land (d  < b  \implies \Gpost\rN{s}{3}) )=   &  \\
& \Arglaw{\lAiB}\\
\t1  ((d  < b  \land\Gpost\rN{s}{3})  \land  (r \neq 0 \lor \Gpost\rN{s}{3}))  \lor   ((d \geq b  \land \Gpost  \land  (r \neq 0 \lor \Gpost\rN{s}{3})) \lor\\
\t1   ((r = 0 \land \Gpost\rN{s}{3}) \land (d  \geq b  \lor \Gpost\rN{s}{3})) \lor ((r \neq 0 \land \Gpost) \land (d  \geq b  \lor \Gpost\rN{s}{3}) )=   &  \\
\t1  (d  < b   \land  r \neq 0  \land\Gpost\rN{s}{3} )\lor\\
\t1  (d  < b  \land\Gpost\rN{s}{3}  \land  \Gpost\rN{s}{3})  \lor &\Arglaw{\Gpost\rN{s}{3} \land \Gpost= \Gpost\rN{s}{3}}\\
\t1  (d \geq b \land  r \neq 0 \land \Gpost  ) \lor\\
\t1  (d \geq b  \land \Gpost  \land \Gpost\rN{s}{3}) \lor\\
\t1   ( d  \geq b \land r = 0 \land \Gpost\rN{s}{3}) )  \lor\\
\t1   (r = 0 \land \Gpost\rN{s}{3}) \land  \Gpost\rN{s}{3}) \lor & \Arglaw{\Gpost\rN{s}{3} \land \Gpost= \Gpost\rN{s}{3}}\\
\t1  ( d  \geq b \land r \neq 0 \land \Gpost )   \lor\\
\t1  (r \neq 0 \land  \Gpost\rN{s}{3}) =  &  \\
\t1  (d  < b   \land  r \neq 0  \land\Gpost\rN{s}{3} )\lor & \\
\t1  (d  < b  \land\Gpost\rN{s}{3}  \lor\\
\t1  (d \geq b  \land \Gpost\rN{s}{3}) \lor & \\
\t1  (d \geq b \land  r \neq 0 \land \Gpost  ) \lor\\
\t1   ( d  \geq b \land r = 0 \land \Gpost\rN{s}{3}) )  \lor\\
\t1   (r = 0 \land \Gpost\rN{s}{3})  \lor \\
\t1  (r \neq 0 \land  \Gpost\rN{s}{3})   &  \\
 \end{argue}
 

Assuming $\pra:\ppra$, $\prb:\pprb$, $\prc:\pprc$, and $\prd:\pprd$, and using multiplication instead of conjunction and addition instead of disjunction, the above predicate can be rewritten as:

\begin{argue}
 \t1 \pra\prbn\prd +
 \pra\prd +
  \pran\prd + 
 \pran\prbn\prc +
 \pran\prb \prd +
 \prb\prd+ 
 \prbn\prd=
\end{argue}

\begin{figure}[h]
\centering
\scriptsize
\roundbox{
\begin{minipage}{7cm}

\karnaughmap{4}{$f$}%
{{$d$}{$b$}{$c$}{$a$}}%
{%
00100000%
11111111%
}%
{%
\textcolor{Red}{
\put(2,1){\oval(3.9,1.9)[]}}
\textcolor{Blue}{
\put(.35,2){\oval(.9,1.9)[]}}
}
\end{minipage}
}
\normalsize
\caption{\label{karnobfig}
Karno Map for $ \pra\prbn\prd + \pra\prd +  \pran\prd + \pran\prbn\prc +\pran\prb \prd + \prb\prd+ \prbn\prd$ }
\end{figure}

Using Karno Map(Figure \ref{karnobfig}), this predicate can be simplified as:
\begin{argue}
\t1 \prd +\pran\prbn\prc 
\end{argue}

  Which is:
\begin{argue}
 \t1 \pprd \lor (\ppran \land \pprbn \land \pprc) 
 \end{argue}
 
 So:
\begin{argue}
\Infections =\Infection & \Argcalc{\uparrow}\\
(\ppra \land \pprd) \lor (\ppran \land \pprc)  \lor (\pprb \land \pprd)  \lor (\pprbn \land \pprc) \neq\\
 \pprd \lor (\ppran \land \pprbn \land \pprc) 
\end{argue}

Compairing two Karno Maps, the differences are:
$\pra\prbn\prc\prdn,\pran\prb\prc\prdn, \pran\prbn\prcn\prd$.
Which means:\\
$1)\ppra \land \pprbn \land \pprc \land \pprdn,\\
2)\ppran \land \pprb \land \pprc \land \pprdn, \\
3)\ppran \land \pprbn \land \pprcn \land \pprd$.

 So in other cases than above the infection is $\falsest$. In these three cases, cases 1 and 2 are not feasible since $r$ must be zero. Case 2 shows that when $a > 0$ the process enters the loop and if the last digit($d$) of input($a$)  is greater than base($b$) then the result of executing $\STu$  is different from  the executing  $\STm$, as a result the infection is $\truest$. So the infection condition is : $ a > 0 \land (a-\floor{\frac{a}{10}}) \geq b$

    \begin{argue}
    \underline{{\bf Propagation}}: & \Argdef{ppgdef}\\   
 \Propagations(\STu,\STm)=  \Propagation &  \Argdef{dscwp}\\
 \wpre{\STu, \wpre{\PRa,\Gpost}} \neq \wpre{\STm, \wpre{\PRa,\Gpost}}\\
\end{argue}
\begin{argue}
\wpre{\PRa,\Gpost}=  & \Argdef{chkdigdefb}\\
 \t1 \wpre{\ChkdigPRa,\Gpost}= & \Argdef{drtwp}\\
\t1  \Gpost\rn{\AMDu}{s} = & \\
\end{argue}
\begin{argue}
 \wpre{\STu, \wpre{\PRa,\Gpost}} = &\Argcalc{\uparrow} ,\Argdef{chkdigdefb}\\ 
\t1   \wpre{\ChkdigSTu, \Gpost\rn{\AMDu}{s} } =\\
 & \Argcalc{Infection}\\
\t1  ( (\pra \land \prd) \lor (\pran \land \prc)  \lor (\prb \land \prd)  \lor (\prbn \land \prc)) \rN{\Gpost}{\Gpost\rn{\AMDu}{s}} = & \\
\t1   (\pra \land \Gpost\rn{\AMDu}{3}) \lor (\pran \land\Gpost\rn{\AMDu}{s} )  \lor\\
\t1  (\prb \land \Gpost\rn{\AMDu}{3})  \lor (\prbn \land \Gpost\rn{\AMDu}{s})   & \\
\end{argue}
\begin{argue}
 \wpre{\STm, \wpre{\PRa,\Gpost}} = & \Argcalc{\uparrow} ,\Argdef{chkdigdefb}\\ 
\t1   \wpre{\ChkdigSTm, \Gpost\rn{\AMDmb}{s} } = \\
& \Argcalc{Infection}\\
\t1 ( \prd \lor (\pran \land \prbn \land \prc) )\rN{\Gpost}{\Gpost\rn{\AMDu}{s}}= &\\
\t1  (\Gpost\rn{\AMDu}{3}) \lor (\pran \land \prbn \land \Gpost\rn{\AMDu}{s})  &\\
\end{argue}

  \begin{argue}
 \Propagations= \Propagation \\
\t1   (\pra \land \Gpost\rn{\AMDu}{3}) \lor (\pran \land\Gpost\rn{\AMDu}{s} )  \lor\\
\t1  (\prb \land \Gpost\rn{\AMDu}{3})  \lor (\prbn \land \Gpost\rn{\AMDu}{s})   \neq \\
 \t1  (\Gpost\rn{\AMDu}{3}) \lor (\pran \land \prbn \land \Gpost\rn{\AMDu}{s}) 
  \end{argue}
  This shows that  the propagation condition is $\truest$.

  \begin{argue}
   \underline{{\bf Full~ Test~ Specification}}: &  \Argdef{ftsdefrip}\\
 \FTSrips(\STu,\STm) = \\
 \FTSrip = & \Argcalc{\uparrow}\\
\t2  (b > 1 \land b \leq 10 \land a \geq 0) \land ( r \leq  0 \lor d \geq b)  \land \\
\t2 a > 0 \land (a-\floor{\frac{a}{10}}) \geq b  \land \\
\t2 \truest =\\
\t2  (b > 1 \land b \leq 10 \land a \geq 0) \land ( r \leq  0 \lor d \geq b)  \land \\
\t2 a > 0 \land (a-\floor{\frac{a}{10}}) \geq b 
 \end{argue}
 The full test specification shows that the test case which can kill the mutant($\AMDmb$) strongly is the one with base greater than 1 and less than or equal to 10($b > 1 \land b \leq 10$) and the last digit($d$) of input($a$) must be greater than or equal to base($b$). In this case the result of $\AMDu$  is 2 but the result of $\AMDmb$ is 3.

%% file: zp0406.tex
\subsection{Search Program: Guarded Command Modification}
\begin{EExample}
\label{exampsearch}
Calculate the Reachability, Infection, Propagation and Full Test Specification conditions for the mutant of {\tt Search} program. The original code in which the mutated statement is marked by prime is as follow:

\psearchvb

Solution:
The complete test
specification to kill this mutant strongly is:
\begin{argue}
Reachability: b.length>0\\
Infection:(x = b[i] \neq x <= b[i]) = (x \notin b \land \exists i @ x<b[i])  \lor (x = b[i] \land \exists j | j < i @ x<b[j])\\
Propagation:x \notin b \\
Full~ Test~ Specification:(b.length>0) \land ((x \notin b \land \exists i @ x<b[i])  \lor (x = b[i] \land \exists j | j < i @ x<b[j])) \land (x \notin b)\\
\t3 (b.length>0) \land (x \notin b \land \exists i @ x<b[i])
\end{argue}
\end{EExample}
\subsubsection{Definitions}

The translation of the program discussed in Example \ref{exampsearch} and its mutated version  to Dijkestra Guarded Commands are as follow:
 \renewcommand{\AMD}{Search}
\renewcommand{\AMDu}{\undertest{\AMD}}
\renewcommand{\AMDm}{\mutated{\AMD}}
\[\AMDu\defs \begin{array}{l} \psearch \end{array} \]
\[\AMDm \defs\begin{array}{l} \psearchm \end{array} \]

\newcommand{\searchPRb}{\left\{\begin{array}{l} \psearchb \end{array}\right\}}

\newcommand{\searchSTu}{\left\{\begin{array}{l} \psearchstu \end{array}\right\}}
\newcommand{\searchSTuif}{\left\{\begin{array}{l} \psearchstuif \end{array}\right\}}
\newcommand{\searchPRjb}{\psearchprjb}
\newcommand{\searchGj}{\psearchgj }
\newcommand{\searchGdoj}{\searchGj }
\newcommand{\searchPRjSTu}{\psearchprjstu}
\newcommand{\searchPRja}{\psearchprja}
\newcommand{\searchPRj}{
\left\{\begin{array}{l}
 \searchPRjSTu \compose \\
 \searchPRja
 \end{array}\right\}}
\newcommand{\searchPRa}{\psearcha}
\newcommand{\searchSTm}{\left\{\begin{array}{l} \psearchstm \end{array}\right\} }
\newcommand{\searchSTmif}{\left\{\begin{array}{l} \psearchstmif \end{array}\right\} }
\newcommand{\searchPRjSTm}{\psearchprjstm}
\newcommand{\searchGdo}{\psearchgj}
\newcommand{\searchIdo}{(\forall k:\nat | 0\leq k < i @ b~k \neq x)}
\newcommand{\searchVdo}{(l-i)}
\newcommand{\searchVardo}{i,\AMD}
\newcommand{\searchVarfdo}{z}

  The underlined statement is the mutated statement. This is an example  of a guarded command modification. To calculate the RIP conditions,  the above programs and what are given in Definitions \ref{putdefn} and  \ref{mutdefn} must be compared. So:\\
\begin{Definition} 
\label{dsearchdef}
\begin{syntax}
\PRb & : &  \searchPRb  \\
\STu & : & \searchSTu   \\
\PRjb &:& \searchPRjb\\
\STju & : & \searchPRjSTu   \\
\PRja &:& \searchPRja\\
\PRa & : & \searchPRa \\
\STm & : &  \searchSTm\\
\STjm & : &  \searchPRjSTm\\
\Gj  &:&   \searchGj\\
\Gdo   &:& \searchGdo\\
\Ido   &:& \searchIdo\\
\Ido \land \Gdo &:& (\searchIdo \land  \searchGdo)\\
\Ido \land \Gj &:& (\searchIdo \land  \searchGj)\\
\Vdo   &:& \searchVdo\\
\end{syntax}
\end{Definition}

\subsubsection{Checking \LWDC\, Conditions}
First, we must check the \LWDC\, conditions given in Definitions \ref{ddowpba} and \ref{ddowpc}, and knowing that we have only one guard, which means $i=1$.
\begin{argue}
\LWDC  & \Argdef{dsearchdef}\\
\forall \searchVardo @  (\Ido \land \Gj \implies \wpre{\searchPRj,\Ido}) \land \\
\t1  (\Ido \land \Gj \implies   \wpre{\searchVarfdo:=\searchVdo \compose \searchPRj, \searchVdo < \searchVarfdo}))  \land \\
\t1 (\Ido \land \Gdo \implies \searchVdo >0)= & \Argdef{dscwp}\\
\forall \searchVardo @  (\Ido \land \Gj \implies \wpre{\searchPRjSTu, \wpre{\searchPRja,\Ido}}) \land  \\
\t1  (\Ido \land \Gj \implies   \wpre{\searchVarfdo:=\searchVdo, \wpre{\searchPRjSTu, \wpre{\searchPRja, \searchVdo < \searchVarfdo}}}))  \land \\
\t1 (\Ido \land \Gdo \implies \searchVdo >0)= & \Argdef{daswp}\\
\forall \searchVardo @  (\Ido \land \Gj \implies \wpre{\searchPRjSTu, \Ido\rN{i}{i+1}}) \land \\
\t1  (\Ido \land \Gj \implies   \wpre{\searchVarfdo:=\searchVdo, \wpre{\searchPRjSTu,  (\searchVdo < \searchVarfdo)\rN{i}{i+1}}}))  \land \\
\t1 (\Ido \land \Gdo \implies \searchVdo >0)= & \Argdef{difwp} \\
\forall \searchVardo @  (\Ido \land \Gj \implies ((x=b~i) \lor (x \neq b~i)) \land & \Arglaw{\lAonA}\\
\t2  ((x=b~i) \implies \wpre{\returnst(1), \Ido\rN{i}{i+1}}) \land & \Argdef{drtwp}\\
\t2  ((x \neq b~i) \implies \wpre{\skipst, \Ido\rN{i}{i+1}}) \land  & \Argdef{dskwp}\\
\t1  (\Ido \land \Gj \implies   \wpre{\searchVarfdo:=\searchVdo, ((x=b~i) \lor (x \neq b~i)) \land & \Arglaw{\lAonA}\\
\t2  ((x=b~i) \implies \wpre{\returnst(1),  (\searchVdo < \searchVarfdo)\rN{i}{i+1}}) \land & \Argdef{drtwp} \\
\t2  ((x \neq b~i) \implies \wpre{\skipst,  (\searchVdo < \searchVarfdo)\rN{i}{i+1}}}) \land & \Argdef{dskwp} \\
\t1 (\Ido \land \Gdo \implies \searchVdo >0)= & \\
\forall \searchVardo @  (\Ido \land \Gj \implies \truest \land\\
\t2  ((x=b~i) \implies  \Ido\rN{i}{i+1}\rN{\AMD}{1}) \land \\
\t2  ((x \neq b~i) \implies  \Ido\rN{i}{i+1}) \land \\
\t1  (\Ido \land \Gj \implies   \wpre{\searchVarfdo:=\searchVdo, \truest \land\\
\t2  ((x=b~i) \implies (\searchVdo < \searchVarfdo)\rN{i}{i+1}\rN{\AMD}{1}) \land \\
\t2  ((x \neq b~i) \implies (\searchVdo < \searchVarfdo)\rN{i}{i+1}}) \land \\
\t1 (\Ido \land \Gdo \implies \searchVdo >0)= & \\
\forall \searchVardo @  (\Ido \land \Gj \implies \\
\t2  ((x=b~i) \implies  \Ido\rN{i}{i+1}\rN{\AMD}{1}) \land \\
\t2  ((x \neq b~i) \implies  \Ido\rN{i}{i+1}) \land \\
\t1  (\Ido \land \Gj \implies   \wpre{\searchVarfdo:=\searchVdo, & \Argdef{daswp}\\
\t2  ((x=b~i) \implies (\searchVdo < \searchVarfdo)\rN{i}{i+1}\rN{\AMD}{1}) \land \\
\t2  ((x \neq b~i) \implies (\searchVdo < \searchVarfdo)\rN{i}{i+1}}) \land \\
\t1 (\Ido \land \Gdo \implies \searchVdo >0)= & \\
\forall \searchVardo @  (\Ido \land \Gj \implies \\
\t2  ((x=b~i) \implies  \Ido\rN{i}{i+1}\rN{\AMD}{1}) \land \\
\t2  ((x \neq b~i) \implies  \Ido\rN{i}{i+1}) \land \\
\t1  (\Ido \land \Gj \implies   & \\
\t2  (((x=b~i) \implies (\searchVdo < \searchVarfdo)\rN{i}{i+1}\rN{\AMD}{1}) \land \\
\t2  ((x \neq b~i) \implies (\searchVdo < \searchVarfdo)\rN{i}{i+1}) \rN{\searchVarfdo}{\searchVdo}) \land \\
\t1 (\Ido \land \Gdo \implies \searchVdo >0)= & \\
\forall \searchVardo @  (\searchIdo \land \searchGj \implies \\
\t2  ((x=b~i) \implies  \searchIdo\rN{i}{i+1}\rN{\AMD}{1}) \land \\
\t2  ((x \neq b~i) \implies  \searchIdo\rN{i}{i+1}) \land \\
\t1  (\searchIdo \land \searchGj \implies   & \\
\t2  (((x=b~i) \implies (\searchVdo < \searchVarfdo)\rN{i}{i+1}\rN{\AMD}{1}) \land \\
\t2  ((x \neq b~i) \implies (\searchVdo < \searchVarfdo)\rN{i}{i+1})) \rN{\searchVarfdo}{\searchVdo} \land \\
\t1 (\searchIdo \land \searchGdo \implies \searchVdo >0)= & \\
\forall \searchVardo @  (\searchIdo \land \searchGj \implies \\
\t2  ((x=b~i) \implies  (\forall k:\nat | 0\leq k < (i+1) @ b~k \neq x)\rN{\AMD}{1}) \land \\
\t2  ((x \neq b~i) \implies  (\forall k:\nat | 0\leq k < (i+1) @ b~k \neq x)) \land \\
\t1  (\searchIdo \land \searchGj \implies   & \\
\t2  (((x=b~i) \implies (l-(i+1) < \searchVarfdo)\rN{\AMD}{1}) \land \\
\t2  ((x \neq b~i) \implies (l-(i+1) < \searchVarfdo))) \rN{\searchVarfdo}{\searchVdo} \land \\
\t1 (\searchIdo \land \searchGdo \implies \searchVdo >0)= & \\
\forall \searchVardo @  (\searchIdo \land \searchGj \implies \\
\t2  ((x=b~i) \implies  (\forall k:\nat | 0\leq k < (i+1) @ b~k \neq x)\rN{\AMD}{1}) \land \\
\t2  ((x \neq b~i) \implies  (\forall k:\nat | 0\leq k < (i+1) @ b~k \neq x)) \land \\
\t1  (\searchIdo \land \searchGj \implies   & \\
\t2  (((x=b~i) \implies (l-(i+1) < \searchVdo)\rN{\AMD}{1}) \land \\
\t2  ((x \neq b~i) \implies (l-(i+1) < \searchVdo))) \land \\
\t1 (\searchIdo \land \searchGdo \implies \searchVdo >0)= & \\
\t1 \truest
\end{argue}
\subsubsection{RIP Calculations}

\begin{argue}
\underline{{\bf Reachability}}: &\Argdef{rchdefn}\\
\Reachabilityg(\STju)=  \Reachabilityn   & \Argdef{dsearchdef}\\
\Reachabilityg(\searchPRjSTu)=     & \\
\t1 \reachc{\searchPRb}  \land  \searchGj \land  \reachc{\searchPRjb} = & \Argdef{dscrc}\\
\t2 \reachc{i := 0}  \land  \reachc{l := b.length } \land & \Argdef{dasrc}\\ 
\t2 \searchGj \land  & \Argcalc{i=0 \land l := b.length} \\
 \t2 \reachc{\nullst}= & \Argdef{dnlrc}\\
\t2 \truest  \land  \truest \land \\
\t2 0 < b.length \land \\   
\t2 \truest =\\
\t2 0 < b.length
\end{argue}
The reachability condition shows that $b$ must not be $\nullst$ or empty .

\begin{argue}
\underline{{\bf Infection}}: &  \Argdef{infdefn}\\
\Infectiong(\STju,\STjm)= \Infectionn & \\
\wpre{\STju,\Gpost}=  & \Argdef{dsearchdef}\\
\t1  \wpre{\searchPRjSTu,\Gpost}= &  \Argdef{difwp} \\
\t1   ((x=b~i) \lor (x\neq b~i)) \land  & \Arglaw{\lAonA}\\
\t1 ((x=b~i) \implies \wpre{\returnst(1),\Gpost}) \land & \Argdef{drtwp} \\
\t1 ( (x\neq b~i) \implies \wpre{ \skipst,\Gpost})= & \Argdef{dskwp}\\
\t1  \truest \land  ((x=b~i) \implies \Gpost\rN{\AMDu}{1}) \land ( (x\neq b~i) \implies \Gpost)= & \\
\t1  ((x=b~i) \implies \Gpost\rN{\AMDu}{1}) \land ( (x\neq b~i) \implies \Gpost)= &  \Arglaw{\laneb}\\
\t1  ((x=b~i) \implies \Gpost\rN{\AMDu}{1}) \land ( ((x< b~i) \lor (x > b~i)) \implies \Gpost)= & \\
 & \Arglaw{\lAoBiC}\\
\t1  ((x=b~i) \implies \Gpost\rN{\AMDu}{1}) \land ( (x< b~i)  \implies \Gpost) \land  ( (x > b~i) \implies \Gpost)& \\
 \end{argue}
 
 \begin{argue}
 \wpre{\STjm,\Gpost}=  & \Argdef{dsearchdef}\\
\t1  \wpre{\searchPRjSTm,\Gpost}= &  \Argdef{daswp} \\
\t1   ((x\leq b~i) \lor (x > b~i)) \land  &  \Arglaw{\lAonA}\\
\t1 ( (x\leq b~i) \implies \wpre{\returnst(1),\Gpost})\land  & \Argdef{drtwp}\\
\t1 ((x > b~i) \implies \wpre{\skipst,\Gpost}) = &  \Argdef{dskwp} \\
\t1   \truest \land   ( (x\leq b~i) \implies \Gpost\rN{\AMDm}{1})\land  ((x > b~i) \implies \Gpost) = &  \\
\t1   ( (x\leq b~i) \implies \Gpost\rN{\AMDm}{1})\land  ((x > b~i) \implies \Gpost)=  &  \Arglaw{\laleb}\\
\t1   ( ((x= b~i) \lor (x < b~i)) \implies \Gpost\rN{\AMDm}{1})\land  ((x > b~i) \implies \Gpost)=  &  \\
& \Arglaw{\lAoBiC}\\
\t1   ( (x = b~i) \implies \Gpost\rN{\AMDm}{1})\land ((x< b~i)  \implies \Gpost\rN{\AMDm}{1}) \land  ((x > b~i) \implies \Gpost)\\
\end{argue}
\begin{argue}
\underline{{\bf Infection}}: &  \Argdef{infdefn}\\
\Infectiong(\STju,\STjm)= \Infectionn &\Argcalc{\uparrow} \\
\t1  ((x=b~i) \implies \Gpost\rN{\AMDu}{1}) \land ( (x< b~i)  \implies \Gpost) \land  ( (x > b~i) \implies \Gpost) \neq \\
\t1  ( (x = b~i) \implies \Gpost\rN{\AMDm}{1})\land ((x< b~i)  \implies \Gpost\rN{\AMDm}{1}) \land  ((x > b~i) \implies \Gpost) \\
 \end{argue}

The the difference is when $(x< b~i)$. It means before finding $x$ in $b$ if we reach to an $i$ so that $x < b~i$ the infection will occur.
So the Infection condition can be formalized as: 
\[ ((\exists i @ x= b~i) \land  (\exists j | j <i @ x < b ~j)) \lor ((x \notin b) \land \exists i @ x < b~i)\]

    \begin{argue}
    \underline{{\bf Propagation}}: & \Argdef{ppgdefndo}\\   
 \Propagationgdo(\STju,\STjm)=  \Propagationndo  &   \Argdef{dscwp}\\
\wpre{\STu, \wpre{\PRa,\Gpost}}  \neq \wpre{ \STm ,\wpre{ \PRa,\Gpost}}=\\
\end{argue}
\begin{argue}
\wpre{\PRa,\Gpost}=   \underbrace{\wpre{\searchPRa,\Gpost}}_{\Gpostb} & \Argdef{dsearchdef}\\
\end{argue}
\renewcommand{\AGpostb}{\wpre{\searchPRa,\Gpost}}
\begin{argue}
\wpre{\STu, \wpre{\PRa,\Gpost}}= & \Argdef{dsearchdef},\Argcalc{\uparrow}\\
\t1 \wpre{\searchSTu,\Gpostb } = & \Argdef{ddowp},\Argcalc{k=1}\\
 \t1 \DHp{1}{\Gpostb} = & \Argdef{ddowp}\\
\t1 \wpre{\searchSTuif, \DHp{0}{\Gpostb}}  =& \Argdef{difwp}\\
\t1 (\searchGdo \lor \lnot\searchGdo) \land & \Arglaw{\lAonA} \\
\t1 \searchGdo \implies  \wpre{\searchPRjb\compose \searchPRjSTu\compose\searchPRja,\DHp{0}{\Gpostb} } \land & \Argdef{dscwp}\\
\t1 \lnot \searchGdo \implies  \wpre{\skipst,\DHp{0}{\Gpostb} }= & \Argdef{dskwp}\\
\t1 \truest \land \\
\t1 \searchGdo \implies\\
  \wpre{\searchPRjb, \wpre{ \searchPRjSTu, \wpre{\searchPRja,\DHp{0}{\Gpostb} }}} \land & \Argdef{daswp}\\
\t1 \lnot \searchGdo \implies  \DHp{0}{\Gpostb} = & \\
\t1 \searchGdo \implies  \wpre{\searchPRjb, \wpre{ \searchPRjSTu,\DHp{0}{\Gpostb}\rN{i}{i+1 }}} \land  & \Argdef{difwp}\\
\t1 \lnot \searchGdo \implies  \DHp{0}{\Gpostb} = & \\
\t1 \searchGdo \implies  \wpre{\searchPRjb,  ((x=b~i) \lor (x\neq b~i)) \land & \Arglaw{\lAonA}\\
\t2  ((x=b~i) \implies \wpre{\returnst(1), \DHp{0}{\Gpostb}\rN{i}{i+1 }}) \land  & \Argdef{drtwp}\\
\t2  ((x\neq b~i) \implies \wpre{\skipst,\DHp{0}{\Gpostb}\rN{i}{i+1 }}} \land  & \Argdef{dskwp}\\
\t1 \lnot \searchGdo \implies  \DHp{0}{\Gpostb} = & \\
\t1 \searchGdo \implies  \wpre{\searchPRjb, \truest \land & \\
\t2  ((x=b~i) \implies \DHp{0}{\Gpostb} \rN{i}{i+1 }\rN{\AMDu}{1}) \land  & \\
\t2  ((x\neq b~i) \implies \DHp{0}{\Gpostb}\rN{i}{i+1 }} \land  & \\
\t1 \lnot \searchGdo \implies  \DHp{0}{\Gpostb} = &  \\
\end{argue}
We know that:
\begin{argue}
\DHp{0}{\Gpostb}=\Gpostb \land \lnot \searchGdo & \Argdef{ddowp}\\
\Gpostb=\AGpostb & \Argcalc{\uparrow}\\
\AGpostb = \Gpost\rN{\AMDu}{0} & \Argcalc{\uparrow} \\
\DHp{0}{\Gpostb}=  \Gpost\rN{\AMDu}{0}  \land \lnot \searchGdo
\end{argue}
So we have:
\begin{argue}
\wpre{\STu, \wpre{\PRa,\Gpost}}= & \Argcalc{\uparrow}\\
\t1 \searchGdo \implies  \wpre{\searchPRjb, & \\
\t2  ((x=b~i) \implies (\Gpost\rN{\AMDu}{0} \land \lnot \searchGdo) \rN{i}{i+1 }\rN{\AMDu}{1}) \land  & \\
\t2  ((x\neq b~i) \implies  (\Gpost\rN{\AMDu}{0}\land \lnot \searchGdo)\rN{i}{i+1 })} \land  & \\
\t1 \lnot \searchGdo \implies   (\Gpost\rN{\AMDu}{0} \land \lnot \searchGdo) = &  \\
\t1 \searchGdo \implies  \wpre{\searchPRjb, & \\
\t2  ((x=b~i) \implies (\Gpost\rN{\AMDu}{1} \land \lnot ((i+1)<l))) \land  & \\
\t2  ((x\neq b~i) \implies  (\Gpost\rN{\AMDu}{0}\land \lnot ((i+1)<l)))} \land  & \\
\t1 \lnot \searchGdo \implies   (\Gpost\rN{\AMDu}{0} \land \lnot \searchGdo) = & \Argdef{dnlwp},\Arglaw{\laneb} \\
\t1 \searchGdo \implies   ((x=b~i) \implies (\Gpost\rN{\AMDu}{1} \land \lnot ((i+1)<l))) \land \\
\t3  (((x< b~i) \lor (x> b~i)) \implies  (\Gpost\rN{\AMDu}{0}\land \lnot ((i+1)<l))) \land  &  \\
\t1 \lnot \searchGdo \implies   (\Gpost\rN{\AMDu}{0} \land \lnot \searchGdo) = &  \Arglaw{\lAoBiC}\\
\t1 \searchGdo \implies   (((x=b~i) \implies (\Gpost\rN{\AMDu}{1} \land \lnot ((i+1)<l))) \land \\
\t3                                 ((x<b~i) \implies (\Gpost\rN{\AMDu}{0}\land \lnot ((i+1)<l))) \land  &  \\
\t3                                 ((x>b~i) \implies (\Gpost\rN{\AMDu}{0}\land \lnot ((i+1)<l)))) \land  &  \\
\t1 \lnot \searchGdo \implies   (\Gpost\rN{\AMDu}{0} \land \lnot \searchGdo)  &  \\
\end{argue}
\begin{argue}
\wpre{\STm, \wpre{\PRa,\Gpost}}= & \Argdef{dsearchdef},\Argcalc{\uparrow}\\
\t1 \wpre{\searchSTm,\Gpostb } = & \Argdef{ddowp},\Argcalc{k=1}\\
 \t1 \DHp{1}{\Gpostb} = & \Argdef{ddowp}\\
\t1 \wpre{\searchSTmif, \DHp{0}{\Gpostb}}  =& \Argdef{difwp}\\
\t1 (\searchGdo \lor \lnot\searchGdo) \land & \Arglaw{\lAonA} \\
\t1 \searchGdo \implies  \wpre{\searchPRjb\compose \searchPRjSTm\compose\searchPRja,\DHp{0}{\Gpostb} } \land & \Argdef{dscwp}\\
\t1 \lnot \searchGdo \implies  \wpre{\skipst,\DHp{0}{\Gpostb} }= & \Argdef{dskwp}\\
\t1 \truest \land \\
\t1 \searchGdo \implies\\
  \wpre{\searchPRjb, \wpre{ \searchPRjSTm, \wpre{\searchPRja,\DHp{0}{\Gpostb} }}} \land & \Argdef{daswp}\\
\t1 \lnot \searchGdo \implies  \DHp{0}{\Gpostb} = & \\
\t1 \searchGdo \implies  \wpre{\searchPRjb, \wpre{ \searchPRjSTm,\DHp{0}{\Gpostb} \rN{i}{i+1 }}} \land  & \Argdef{difwp}\\
\t1 \lnot \searchGdo \implies  \DHp{0}{\Gpostb} = & \\
\t1 \searchGdo \implies  \wpre{\searchPRjb,  ((x \leq b~i) \lor (x > b~i)) \land & \Arglaw{\lAonA}\\
\t2  ((x\leq b~i) \implies \wpre{\returnst(1), \DHp{0}{\Gpostb}\rN{i}{i+1 }}) \land  & \Argdef{drtwp}\\
\t2  ((x > b~i) \implies \wpre{\skipst,\DHp{0}{\Gpostb}\rN{i}{i+1 }}} \land  & \Argdef{dskwp}\\
\t1 \lnot \searchGdo \implies  \DHp{0}{\Gpostb} = & \\
\t1 \searchGdo \implies  \wpre{\searchPRjb, \truest \land & \\
\t2  ((x \leq b~i) \implies \DHp{0}{\Gpostb}\rN{i}{i+1 }\rN{\AMDm}{1}) \land  & \\
\t2  ((x > b~i) \implies \DHp{0}{\Gpostb}\rN{i}{i+1 }} \land  & \\
\t1 \lnot \searchGdo \implies  \DHp{0}{\Gpostb} = &  \\
\end{argue}
We know that:
\begin{argue}
\DHp{0}{\Gpostb}=\Gpostb \land \lnot \searchGdo & \Argdef{ddowp}\\
\Gpostb=\AGpostb & \Argcalc{\uparrow}\\
\AGpostb = \Gpost\rN{\AMDm}{0} & \Argcalc{\uparrow} \\
\DHp{0}{\Gpostb}=  \Gpost\rN{\AMDm}{0}  \land \lnot \searchGdo
\end{argue}
So we have:
\begin{argue}
\wpre{\STm, \wpre{\PRa,\Gpost}}= & \Argcalc{\uparrow}\\
\t1 \searchGdo \implies  \wpre{\searchPRjb, & \\
\t2  ((x \leq b~i) \implies (\Gpost\rN{\AMDm}{0} \land \lnot \searchGdo)\rN{i}{i+1 }\rN{\AMDm}{1}) \land  & \\
\t2  ((x > b~i) \implies  (\Gpost\rN{\AMDm}{0}\land \lnot \searchGdo)\rN{i}{i+1 })} \land  & \\
\t1 \lnot \searchGdo \implies   (\Gpost\rN{\AMDm}{0} \land \lnot \searchGdo) = &  \\
\t1 \searchGdo \implies  \wpre{\searchPRjb, & \\
\t2  ((x \leq b~i) \implies (\Gpost\rN{\AMDm}{1} \land \lnot ((i+1)<l))) \land  & \\
\t2  ((x >  b~i) \implies  (\Gpost\rN{\AMDm}{0}\land \lnot ((i+1)<l)))} \land  & \\
\t1 \lnot \searchGdo \implies   (\Gpost\rN{\AMDm}{0} \land \lnot \searchGdo) = & \Argdef{dnlwp},\Arglaw{\laleb} \\
\t1 \searchGdo \implies   (((x= b~i) \lor (x < b~i)) \implies (\Gpost\rN{\AMDu}{1} \land \lnot ((i+1)<l))) \land \\
\t3  (((x> b~i) \implies  (\Gpost\rN{\AMDm}{0}\land \lnot ((i+1)<l))) \land  &  \\
\t1 \lnot \searchGdo \implies   (\Gpost\rN{\AMDm}{0} \land \lnot \searchGdo) = &  \Arglaw{\lAoBiC}\\
\t1 \searchGdo \implies   (((x=b~i) \implies (\Gpost\rN{\AMDm}{1} \land \lnot ((i+1)<l))) \land \\
\t3                                 ((x<b~i) \implies(\Gpost\rN{\AMDm}{1} \land \lnot ((i+1)<l)))  \land  &  \\
\t3                                 ((x>b~i) \implies (\Gpost\rN{\AMDm}{0}\land \lnot ((i+1)<l)))) \land  &  \\
\t1 \lnot \searchGdo \implies   (\Gpost\rN{\AMDu}{0} \land \lnot \searchGdo)  &  \\
\end{argue}
    \begin{argue}
    \underline{{\bf Propagation}}: & \Argdef{ppgdefndo}\\   
 \Propagationgdo(\STju,\STjm)=  \Propagationndo  &   \Argdef{dscwp}\\
\wpre{\STu, \wpre{\PRa,\Gpost}}  \neq \wpre{ \STm ,\wpre{ \PRa,\Gpost}}=\\
\t1 \searchGdo \implies   (((x=b~i) \implies (\Gpost\rN{\AMDu}{1} \land \lnot ((i+1)<l))) \land \\
\t3                                 ((x<b~i) \implies (\Gpost\rN{\AMDu}{0}\land \lnot ((i+1)<l))) \land  &  \\
\t3                                 ((x>b~i) \implies (\Gpost\rN{\AMDu}{0}\land \lnot ((i+1)<l)))) \land  &  \\
\t1 \lnot \searchGdo \implies   (\Gpost\rN{\AMDu}{0} \land \lnot \searchGdo)   \neq \\
\t1 \searchGdo \implies   (((x=b~i) \implies (\Gpost\rN{\AMDm}{1} \land \lnot ((i+1)<l))) \land \\
\t3                                 ((x<b~i) \implies(\Gpost\rN{\AMDm}{1} \land \lnot ((i+1)<l)))  \land  &  \\
\t3                                 ((x>b~i) \implies (\Gpost\rN{\AMDm}{0}\land \lnot ((i+1)<l)))) \land  &  \\
\t1 \lnot \searchGdo \implies   (\Gpost\rN{\AMDu}{0} \land \lnot \searchGdo)   &  \\
\end{argue}

The propagation condition is $\truest$ only when $x<b~i$ and the length of $b$ is $1$, since $k$ was $1$, and it means $x \notin b$.

  \begin{argue}
   \underline{{\bf Full~ Test~ Specification}}: &  \Argdef{ftsdefripndo}\\
 \FTSripgdo(\STju,\STjm) = \\
 \FTSripndo = & \Argcalc{\uparrow}\\
\t2  0 < b.length \land \\
\t2((\exists i @ x= b~i) \land  (\exists j | j <i @ x < b ~j)) \lor ((x \notin b) \land \exists i @ x < b~i)  \land \\
\t2 x \notin b =\\
\t2 0 < b.length \land (\exists i @ x < b~i)  \land x \notin b
 \end{argue}
 The full test specification shows that the test case which can kill the mutant($\AMDm$) strongly is e.g. $b=[5], x=2$

%% file: zp0500.tex
\setcounter{chap}{5}
\section{Conclusions}
In this paper the formal analysis of Reachability, Infection, and Propagation conditions in Murtation testing is investigated. The weakest precondition predicate transformer and the reachability condition generator function are used to calculate these conditions. To formalize the definitions of these conditions, two types of modifications are distinguished, namely,  the {\it statement modification} and the {\it guarded command modification}.  Program templates for the under test  module and its mutated version are specified. Based on these templates the Reachability, Infection, and Propagation conditions are formally specified when   the statement modification  or the guarded command modification are applied. The results of applying the proposed method are promising. The incremental nature of the method helps the analyzer to calculate these conditions from small pices of programs to the composition of two or more statements.